\documentclass[12pt]{article}

\usepackage[font=footnotesize]{caption}
\usepackage{longtable}
\usepackage{tabu}
\usepackage{array}
\usepackage{enumerate}
\usepackage[subrefformat=parens,labelformat=parens]{subfig}
\usepackage{multicol}
\usepackage{multirow}
\usepackage{epsfig}
\usepackage{graphicx}
\usepackage{color}
\usepackage{cite}

\usepackage{amsmath,amsfonts,amssymb}
\usepackage{bbm}

\usepackage[utf8]{inputenc}
\usepackage[english]{babel}
\usepackage{xspace}

\usepackage[normalem]{ulem}

\usepackage[
    pdftitle={F-theory duals of singular heterotic K3 models},
    pdfauthor={Fabian Ruehle},
    pdfsubject={},
    bookmarksopen, bookmarksnumbered, bookmarksopenlevel=2, colorlinks=true, linkcolor=blue, citecolor=green, urlcolor=blue]{hyperref}

\newcommand{\U}[1]{\text{U(#1)}\xspace}
\newcommand{\SU}[1]{\text{SU(#1)}\xspace}
\newcommand{\SO}[1]{\text{SO(#1)}\xspace}
\newcommand{\E}[1]{\ensuremath{\text{E}_{#1}}\xspace}

\newcommand{\rep}[1]{\ensuremath{{\mathbf{#1}}}\xspace}

\newcommand{\tr}[0]{\text{tr}\xspace}
\newcommand{\de}{\text{d}}
\let\ii\i
\renewcommand{\i}{\text{i}\,}

\newcommand{\Z}[1]{\ensuremath{\mathbbm{Z}_{#1}}\xspace}
\renewcommand{\P}[1]{\ensuremath{{\mathbbm{P}^{#1}}}\xspace}
\newcommand{\WP}[2]{\ensuremath{{\mathbbm{P}_{#1}^{#2}}}\xspace}
\newcommand{\F}[1]{\ensuremath{{\mathbbm{F}_{#1}}}\xspace}
\newcommand{\fibii}{\ensuremath{\text{\itshape II}}}
\newcommand{\fibiii}{\ensuremath{\text{\itshape III}}}
\newcommand{\fibiv}{\ensuremath{\text{\itshape IV}}}
\definecolor{mypurple}{rgb}{1,0.05,.8}

\begin{document}


\thispagestyle{empty}
\begin{flushright} DESY-14-072\end{flushright}
\vskip 1 cm
\begin{center}
{
  \Large{\bf F-theory duals of singular heterotic K3 models}
}\\[0pt]
\bigskip
\bigskip
{\bf Christoph~L\"udeling$^{\,a,}$\footnote{E-mail: \texttt{christoph.luedeling@gmail.com}}, Fabian~Ruehle$^{\,b,}$}\footnote{E-mail: \texttt{fabian.ruehle@desy.de}}\\[0pt]
\bigskip
\vspace{0.23cm}
{${}^{a}$ Institut f\"ur Gebirgsmechanik GmbH, Friederikenstra{\ss}e~60,~04279~Leipzig,~Germany}\\[2pt]
{${}^{b}$ Deutsches Elektronen-Synchrotron DESY, Notkestrasse~85,~22607~Hamburg,~Germany}
\bigskip
\end{center}

\begin{abstract}
We study F-theory duals of singular heterotic K3 models that correspond to abelian toroidal orbifolds $T^4/\Z{N}$. While our focus is on the standard embedding, we also comment on models with Wilson lines and more general gauge embeddings. In the process of constructing the duals, we work out a Weierstrass description of the heterotic toroidal orbifold models, which exhibit singularities of Kodaira type $I_0^*$, $\fibiv^*$, $\fibiii^*$, and $\fibii^*$. This construction unveils properties like the instanton number per fixed point and a correlation between the orbifold order and the multiplicities in the Dynkin diagram. The results from the Weierstrass description are then used to restrict the complex structure of the F-theory Calabi--Yau threefold such that the gauge group and the matter spectrum of the heterotic theories are reproduced. We also comment on previous approaches that have been employed to construct the duality and point out the differences to our case. Our results show explicitly how the various orbifold models are connected and described in F-theory.
\end{abstract}

\clearpage
\setcounter{page}{1}
\setcounter{footnote}{0}
\setcounter{tocdepth}{2}
\tableofcontents

\section{Introduction}
\label{sec:Introduction}

Heterotic $\E8\times\E8$ orbifold models \cite{Dixon:1985jw,Dixon:1986jc,Ibanez:1986tp} are popular for string phenomenology and for physics beyond the standard model \cite{Buchmuller:2005jr,Buchmuller:2006ik,Lebedev:2006kn,Lebedev:2008un,Blaszczyk:2009in}. Heterotic toroidal abelian orbifolds are based on tori which are orbifolded by acting with discrete abelian groups $\Z{N}$ or $\Z{N}\times\Z{M}$. This introduces mild singularities at the fixed points, which reduces the amount of parallelizable spinors and can thus lead to models with reduced supersymmetry in lower dimensions. Compatibility with the discrete action fixes the complex structure of the underlying tori (except for the case of $\Z2$). These models are based on free conformal field theories which allow for an exact treatment. Orbifolds are at rather special points in the moduli space of string compactifications which exhibit enhanced symmetry. In particular, the primordial $\E8\times\E8$ gauge group is broken rank-preservingly. Furthermore, there are many discrete symmetries which can help solving problems inherent to most models beyond the Standard Model such as proton decay or the $\mu$-problem \cite{Lee:2010gv}. This makes orbifolds phenomenologically attractive, but also rather special points, apparently isolated from each other and from smooth (supergravity) compactifications. In the last years there has been considerable progress connecting different orbifolds with their smooth supergravity counterparts \cite{Lust:2006zh,Nibbelink:2008tv,Blaszczyk:2011hs}; however, the process has to be carried out for each orbifold separately.

F-Theory \cite{Vafa:1996xn} was introduced roughly ten years later as an approach to constructing string vacua which are connected via a web of dualities to type II and heterotic string theories. While these dualities have been worked out for various dimensions of string compactification spaces, we focus on the case of orbifolds corresponding to singular K3 surfaces on the heterotic side, which correspond to F-theory models on (elliptically fibered) Calabi--Yau (CY) threefolds over a complex two-dimensional base space which is a Hirzebruch surface $\F{m}$ \cite{Morrison:1996na, Morrison:1996pp}. Most commonly, the elliptically fibered CY threefolds are discussed in terms of a Weierstrass model, in which the elliptic fiber is given in terms of an equation of degree 6 in the weighted projective space $\mathbbm{P}^2_{231}$ whose coefficients are sections in the base $\F{m}$. Generically, the elliptic fiber degenerates over codimension one subloci in the base. The resolution of the singularities were classified by Kodaira \cite{Kodaira:1963} in terms of the vanishing orders of the the coefficients $f,g$, and the discriminant $\Delta$ of the Weierstrass equation. The resolution requires the introduction of $\P1$'s whose intersection numbers are those of the negative affine Cartan matrix of the ADE-type Lie algebras. Using F-/M-theory duality, it can be seen that the ADE-type singularities give rise to precisely the same gauge group in F-theory. For this reason, we use the name of the singularity in the Kodaira classification and the name of the resulting gauge algebra somewhat interchangeably. Matter arises in codimension two where the singularity type of the fiber is further enhanced. In  the meantime, there has been quite some progress in string phenomenology based on F~theory, ultimately motivated by the possibility to obtain exceptional groups and more general matter representations than in type IIB intersecting brane models. Similar to intersecting branes, however, much of the phenomenological work discusses local models, and global completions are much harder to construct.

For the F-theory duals of smooth models, there is a general algorithm for the construction \cite{Kumar:2009ac,Taylor:2010wm,Kumar:2010ru}.  In contrast, F-/M-Theory on singular spaces is less well understood \cite{Seiberg:1996vs,deBoer:2001px,Braun:2014oya}. As we shall see, also in our case  the tools from the smooth case cannot be applied directly (cf.\ also \cite{Marquart:2002bz}). Furthermore, as we shall explain, toroidal orbifolds do not have a direct Weierstrass description. To circumvent this problem we will use a method used in \cite{Braun:2009wh} to construct a Weierstrass model and from that the heterotic--F-theory duality.

The rest of the paper is organized as follows: In section \ref{sec:Review} we review heterotic orbifold models, F-theory, and the construction of the duality. In the process, we compare to other methods used to construct the duals in the smooth case. In section \ref{sec:OrbifoldDuals} we construct the F-theory duals of all four $T^4/\Z{N}$ orbifolds in the ``standard embedding'' (which means in the case of orbifolds that the discrete gauge bundle is $\Z{N}$, leading to a commutant of $\E7\times\SU2$ for $N=1$ or $\E7\times\U1$ for $N=3,4,6$, respectively) and comment on cases with more general gauge embeddings. In section \ref{sec:Conclusions} we conclude and present an outlook.


\section{Review of heterotic -- F-theory duality}
\label{sec:Review}

\subsection{6D heterotic orbifold models}
\label{sec:6DHeteroticModels}
We consider heterotic orbifold compactifications to six dimensions on Calabi--Yau manifolds, which are singular limits of K3. We study orbifolds of the type $T^4/\Z{N}$ with $N=2,3,4,6$. The spectra and gauge groups that can be obtained in these models without using Wilson lines have been classified in \cite{Honecker:2006qz}. We collect the spectra of the standard embeddings in table \ref{tab:OrbifoldSpectra}. We also included the spectrum of a $T^4/\Z2$ orbifold where one Wilson line in the first torus has been switched on. This Wilson line has two effects: It breaks the gauge group further down (while preserving the rank) and it projects out some of the matter states which are incompatible with the Wilson line. This leads to the fact that the spectrum at the various orbifold fixed points where the Wilson line acts is different from the spectrum where the Wilson line is trivial.

\begin{figure}[t]
  \subfloat[$T^4/\Z2$ orbifold\label{subfig:T4Z2Classical}]
  {
    \includegraphics[width=0.45\textwidth]{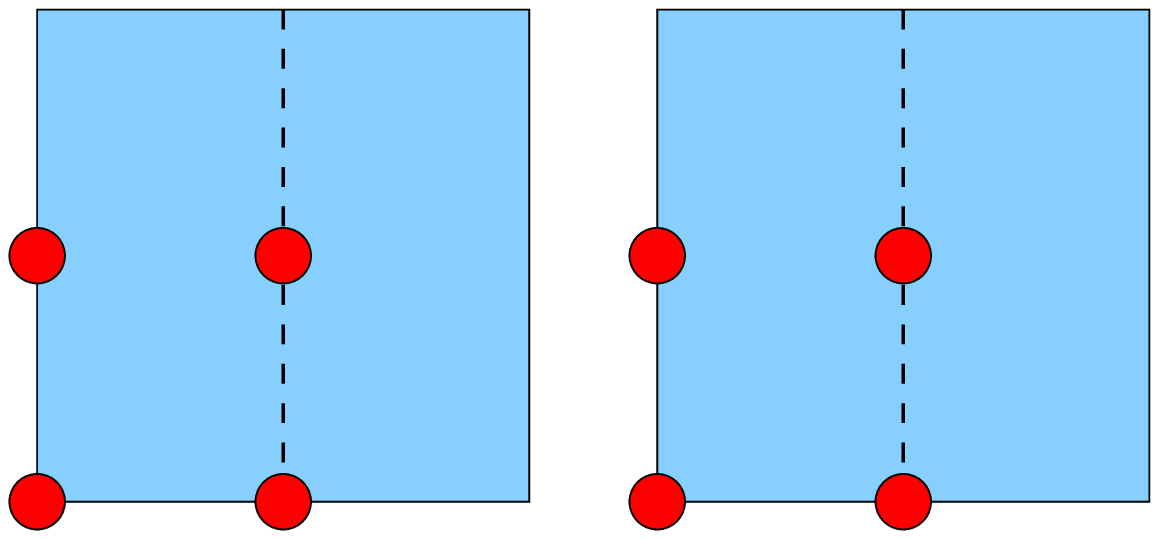}
  }
  \hfill
  \subfloat[Alternative depiction of the $T^4/\Z2$ orbifold\label{subfig:T4Z2Alternative}]
  {
    \includegraphics[width=0.45\textwidth]{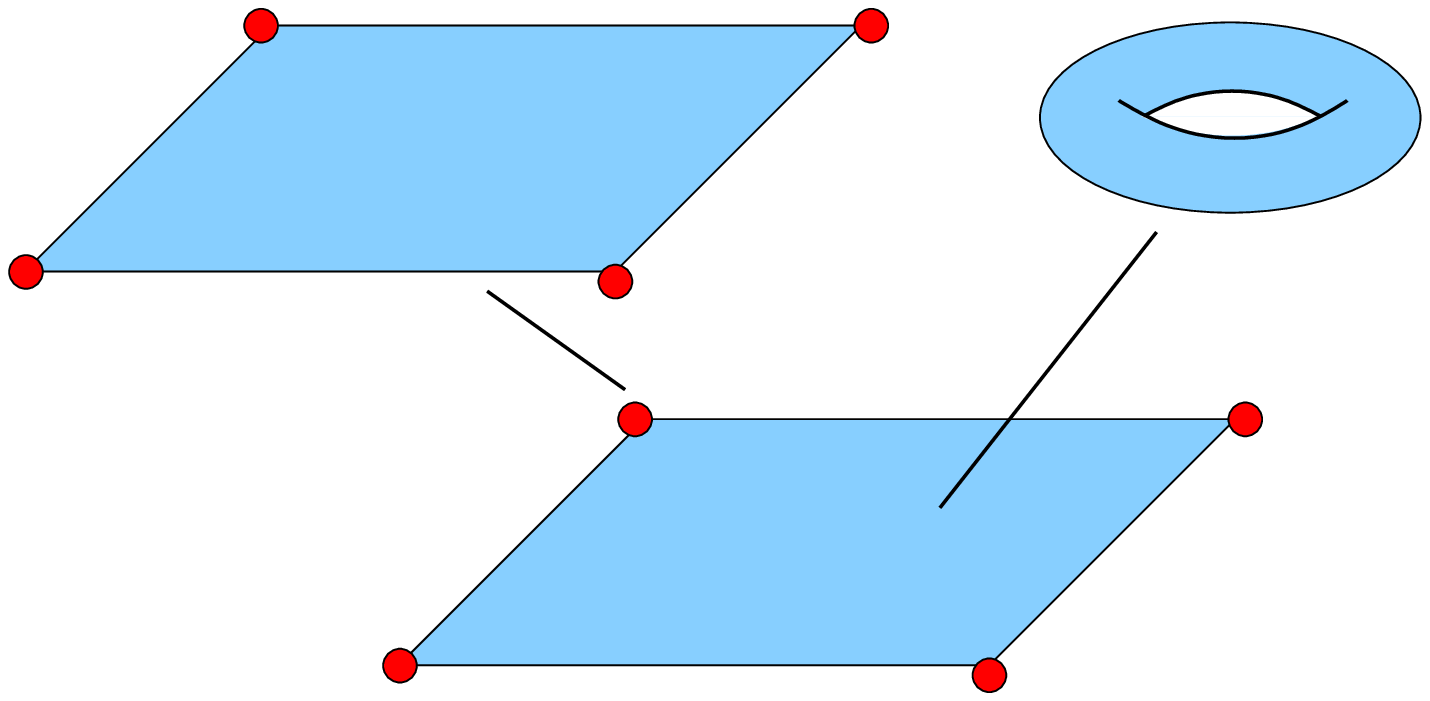}
  }
  \caption{Picture of the $T^4/\Z2$ orbifold. The $4\times4=16$ fixed points are drawn as red dots. The second picture gives an alternative depiction, which will be useful when constructing the F-theory dual later.}
  \label{fig:T4Z2Orbifold}
\end{figure}

For our discussion, we will start with the most simple model, which is the standard embedding of $T^4/\Z{N}$ with gauge group $(\E7\times\SU2)\times\E8$ or $(\E7\times\U1)\times\E8$. The orbifold acts on the two complex torus coordinates $(z_1,z_2)$ as
\begin{align}
 \theta:~(z_1,z_2)\mapsto(e^{2i\pi/N}z_1,e^{-2i\pi/N}z_2)\,.
\end{align}
In order to ensure modular invariance, a twist by $v=\frac1N (1,-1)$ in a $T^4/\Z{N}$ orbifold in standard embedding is accompanied by a shift in the $\E8\times\E8$ gauge degrees of freedom by $V=\frac1N (1,-1,0^6)(0^8)$. 
For the $T^4/\Z2$ model this means that the orbifold acts as a reflection, $(z_1,z_2)\mapsto(-z_1,-z_2)$ and the associated shift breaks the gauge group to $(\E7\times\SU2)\times\E8$. The orbifold action introduces four fixed points in each torus, leading to $16$ $\Z2$ (or $A_1$) orbifold singularities, cf.\ figure \ref{fig:T4Z2Orbifold}. Comparing with table \ref{tab:OrbifoldSpectra}, we see that there are eight $(\rep{56},\rep{1})$. Note, however, that the $\rep{56}$ of $\E7$ is pseudo-real, so there are 16 half-hypers, i.e.\ one per fixed point. Likewise, there are 32 hypermultiplet doublets $(\rep{1},\rep{2})$ which are pseudo-real as well, leading to 4 half-hypers per fixed point. In contrast, the $(\rep{56},\rep{2})$ and the four singlets $(\rep{1},\rep{1})$ in the untwisted sector are not pseudo-real and thus are full hypermultiplets. Also note that there is no matter charged under the second (unbroken) $\E8$. It should be mentioned that the subtlety concerning pseudo-real representations mainly arise in the $T^4/\Z2$ models. The reason is that the commutant of $\Z{N}$ with $\E8$ for $N\neq2$ is $\U1$ rather than $\SU2$. For that reason, the other orbifold standard embeddings come with a \U1 factor, and thus the irreps carrying \U1 charge are complex (but e.g.\ the 5 singlets of the $\Z4$ model are uncharged and correspond to 10 half-hypers at the 6+4 $\Z2$ fixed points occurring in the second twisted sector of $\Z4$). In fact, in all cases where the number of states is half the number of equivalent fixed points the representations are pseudo-real, such that there is one half-hyper per fixed point.

\begin{table}[t]
\tabulinesep=1.2mm
\footnotesize
\centering
 \begin{tabu}{|c|c|c|c|}
  \hline
   Orbifold 			& Gauge Group 					& Untwisted matter 					& Twisted matter\\
   \hline
   \hline
   \multirow{2}{*}{$T^4/\Z2$} 	& $\E7\times\SU2\times\E8$ 			& $(\rep{56},\rep{2})+4(\rep{1},\rep{1})$		& $8(\rep{56},\rep{1};\rep{1})+32(\rep{1},\rep{2})$\\
   \cline{2-4}
				& $\SO{12}\times\SU2^2\times\E8$		& $(\rep{12},\rep{2},\rep{2})+4(\rep{1},\rep{1},\rep{1})$&\parbox{6cm}{$4\,(\rep{32},\rep{1},\rep{1})+4\,(\rep{12},\rep{2},\rep{1})+16\,(\rep{1},\rep{1},\rep{2})+$\\$4\,(\rep{32},\rep{1},\rep{1})+4\,(\rep{12},\rep{1},\rep{2})+16\,(\rep{1},\rep{2},\rep{1})$}\\
   \hline
   $T^4/\Z3$ 			& $\E7\times\U1\times\E8$ 			& $(\rep{56})_1+(\rep{1})_2+2\,(\rep{1})_0$		& $9\,(\rep{56})_{\frac13}+45\,(\rep{1})_{\frac23}+18\,(\rep{1})_{\frac43}$\\
   \hline
   $T^4/\Z4$ 			& $\E7\times\U1\times\E8$ 			& $(\rep{56})_1+2\,(\rep{1})_0$				& \parbox{6cm}{$4\,(\rep{56})_{-\frac12}+8\,(\rep{1})_{\frac32}+24\,(\rep{1})_{\frac12}+$\\$5(\rep{56})_{0}+32\,(\rep{1})_{1}$}\\
   \hline
   $T^4/\Z6$ 			& $\E7\times\U1\times\E8$ 			& $(\rep{56})_1+2\,(\rep{1})_0$				& \parbox{6cm}{$1\,(\rep{56})_{-\frac23}+8\,(\rep{1})_{\frac13}+2\,(\rep{1})_{-\frac53}+$\\$5\,(\rep{56})_{-\frac13}+22\,(\rep{1})_{\frac23}+10\,(\rep{1})_{-\frac43}+$\\$3\,(\rep{56})_{0}+22\,(\rep{1})_{1}$}\\
   \hline
 \end{tabu}
\caption{Spectrum of $T^4/\Z{N}$ orbifold models. Except for the model in the second line, which has one Wilson line, all models are in the orbifold standard embedding. In all models, the second $\E8$ is unbroken, so we omit it in the irreps for brevity. For $\Z4$ and $\Z6$, the $i^\text{th}$ line corresponds to the matter contribution of the $i^\text{th}$ twisted sector.}
\label{tab:OrbifoldSpectra}
\tabulinesep=1.0mm
\end{table}

An advantage of 6D $\mathcal{N}=1$ models is that the chirality of the spinors is fixed: the chirality of the hypermultiplets and tensor multiplets is the same and opposite to the chirality of the vector multiplets. This leads to very stringent anomaly cancellation conditions. In particular, the gravitational anomaly reads
\begin{align}
\label{eq:6DGravAnomaly}
 N_\text{H}-N_\text{V}+29N_\text{T}=273\,,
\end{align}
where $N_\text{H}, N_\text{V}, N_\text{T}$ are the number of hyper-, vector-, and tensor multiplets. Perturbative heterotic string models always have $N_\text{T}=1$, where the scalar of the tensor multiplet is the dilaton. Using \eqref{eq:6DGravAnomaly} and the other conditions ensuring the absence of gauge and mixed anomalies, it can be easily checked that the models in table \ref{tab:OrbifoldSpectra} are anomaly-free. Note that in contrast to 4D, the anomaly \eqref{eq:6DGravAnomaly} also depends on the singlets, i.e.\ it is sensitive to the entire particle spectrum.

It is well-known that these singularities can be resolved, leading to heterotic string models on smooth Calabi--Yaus with vector bundles. In the case of the standard embedding, the vector bundle is $\SU2$ which breaks the $\E8\times\E8$ to the commutant $\E7\times\E8$. The heterotic Bianchi identity for the Kalb--Ramond three-form field strength reads
\begin{align}
\label{eq:BianchiIdentity}
 \de H=\text{ch}_2(TX)-\text{ch}_2(V)\,.
\end{align}
Since the instanton number (i.e.\ the second Chern class) of K3 is $24$, we need to embed a total of $24$ instantons in the gauge bundle to satisfy this identity. In principle, \eqref{eq:BianchiIdentity} is modified in the presence of Neveu--Schwarz five-branes. Including these five-branes will lead to F-theory duals with more than one tensor multiplet\footnote{The scalar in these extra tensor multiplets encodes the position of the five-brane in the M-theory bulk, as explained in section \ref{sec:GeneralInstantonEmbeddings}.}, but we will not say too much about this.

\subsection{Dual F-theory constructions}
\begin{figure}[t]
  \subfloat[CY threefold \label{subfig:CYFib1}]
  {
    \includegraphics[height=4.8cm]{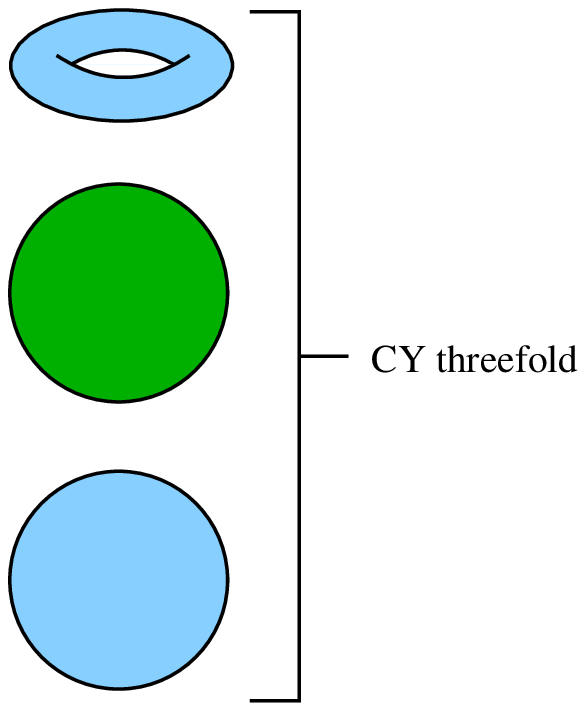}
  }~
  \subfloat[Elliptic fibration \label{subfig:CYFib2}]
  {
    \includegraphics[height=4.8cm]{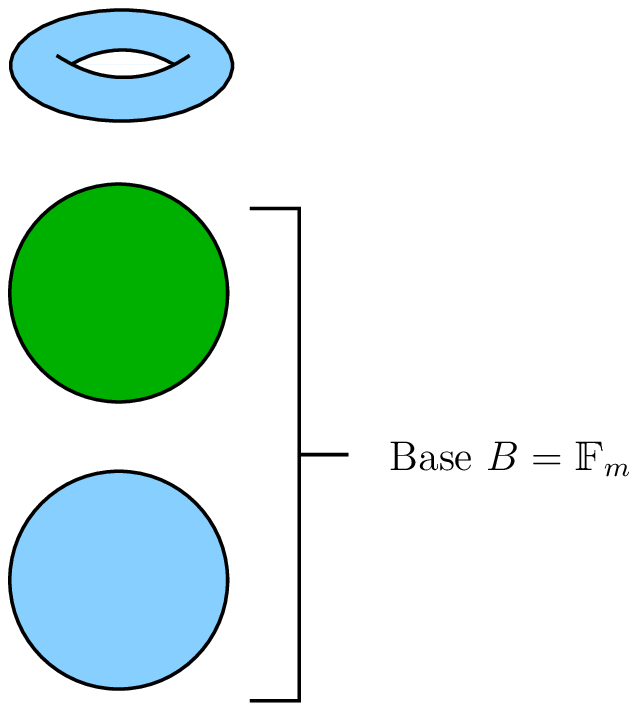}
  }~
  \subfloat[K3 fibration \label{subfig:CYFib3}]
  {
    \includegraphics[height=4.8cm]{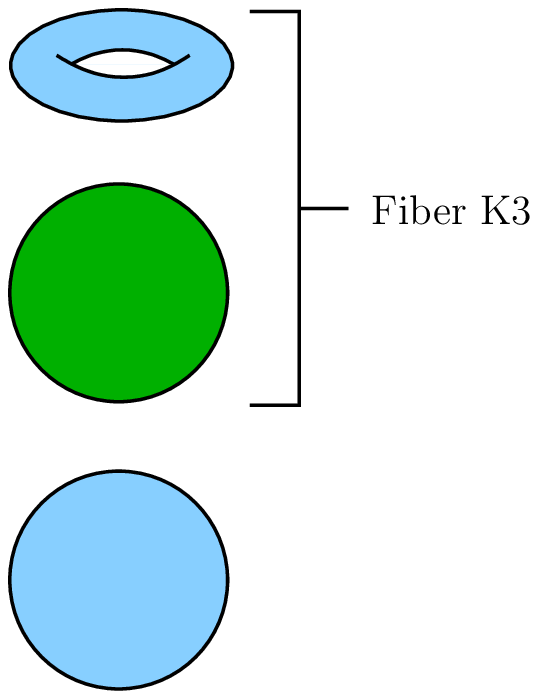}
  }~
  \subfloat[Heterotic dual \label{subfig:CYFib4}]
  {
    \includegraphics[height=4.8cm]{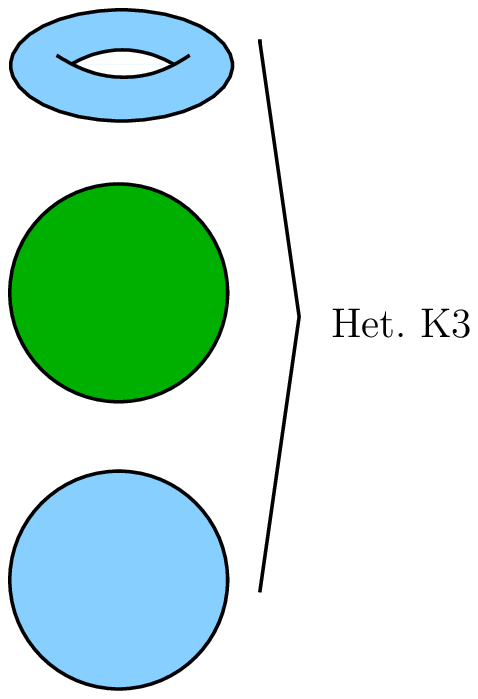}
  }
  \caption{Picture of the CY threefold used in the F-theory construction. It is shown how the CY can be interpreted as: (b) elliptically fibered over $\F{m}$, (c) K3 fibered over $\P1$, and (d) in terms of the heterotic K3 the theory will be dual to.}
  \label{fig:CYFibration}
\end{figure}

F-theory models on Calabi--Yau threefolds $X$ with a heterotic dual show a special fibration structure \cite{Morrison:1996na,Morrison:1996pp}. They are K3 fibrations over $\P1$, where the K3 is itself elliptically fibered over another $\P1$. The base space $B$ of the elliptically fibered CY threefold, i.e.\ the $\P1$ fibration over $\P1$, corresponds to a Hirzebruch surface $\F{m}$. The situation is depicted in figure \ref{fig:CYFibration}. 

Some of the particle content of the underlying theory is fixed by the geometrical data of the CY threefold and the base space \cite{Vafa:1996xn,Morrison:1996pp}. Using, among other things, that in going from the 6D $\mathcal{N}=1$ to the 4D $\mathcal{N}=2$ theory, the 6D vector and tensor multiplets correspond to 4D vector multiplets while the 6D hypermultiplets stay hypermultiplets in 4D, one finds
\begin{align}
\label{eq:FtheoryGeometrySpectrum}
 N_\text{T}=h^{1,1}(B)-1\,,\qquad \text{rk}(V)=h^{1,1}(X)-h^{1,1}(B)-1\,,\qquad N_\text{H}^\text{neutral}=h^{2,1}(X)+1\,.
\end{align}
Here, $\text{rk}(V)$ is the rank of the unbroken 6D gauge group and $N_\text{H}^\text{neutral}$ is the number of uncharged hypermultiplets. As explained above, models with a perturbative heterotic dual have $N_\text{T}=1$. Since $\F{m}$ inherits the two K\"ahler classes of the $\P1$'s, it has $h^{1,1}(B)=2$, which corresponds to $N_\text{T}=1$, as it should be. Furthermore, it has been argued in the smooth case that F-theory with base space $\F{m}$ corresponds to a heterotic $\E8\times\E8$ theory with vector bundles where $12+m$ instantons are embedded in the first \E8 and $12-m$ in the second \E8. All the models in table~\ref{tab:OrbifoldSpectra} have an unbroken hidden \E8 gauge group. For this reason, we will concentrate on $\F{12}$ where all 24 instantons are embedded in the first \E8, leaving the second \E8 intact.

For the description of the elliptic fibered CY threefold we use the Weierstrass model. There the elliptic curve is parametrized as a sextic in the weighted projective space $\WP{231}{2}$,
\begin{align}
\label{eq:WeierstrassShort}
y^2 = x^3 + fxz^4 + gz^6 \, ,
\end{align}
with homogeneous coordinates $x,y,z$. The fibration over the base $\F{12}$ is encoded in $f$ and $g$, which are appropriate sections such that the complete elliptic fibration has a trivial anticanonical bundle. The elliptic fiber degenerates at points where its discriminant
\begin{align}
\label{eq:Discriminant}
\Delta = 4 f^3 + 27 g^2
\end{align}
vanishes. We summarize the scalings\footnote{For $\F{m}$, the $\lambda$-scaling is $m$ for $u$, $2(m+2)$ for $x$, and $3(m+2)$ for $y$.} of the model in table \ref{tab:Scalings}. For the base $\P1$, we denote the homogeneous coordinates by $s,t$ and the scaling by $\mu$. For the fiber $\P1$ we denote the coordinates by $u,v$ and the scaling by $\lambda$; the scaling of the ambient space for the elliptic fiber $\WP{231}{2}$ is denoted by $\nu$.

From the vanishing order of $(f,g,\Delta)$ the gauge group and the matter content can be inferred according to the Kodaira classification \cite{Kodaira:1963}, cf.\ table \ref{tab:Kodaira}. In order to resolve the singular fiber at codimension 1, one glues in extra $\P1$'s. Their intersection numbers with each other and with the original torus are given by the negative of the affine Cartan matrix of the corresponding gauge group.

\begin{table}[t]
\tabulinesep=1.2mm
\centering
 \begin{tabu}{|c||c|c||c|c||c|c|c||c|c|c|}
  \hline
   Scaling  & $s$ & $t$ & $u$ & $v$ & $x$ & $y$ & $z$ & $f$ & $g$ & $\Delta$\\
  \hline
  \hline
  $\lambda$ & $1$ & $1$ & $12$& $0$ & $28$& $42$& $0$ & $56$& $84$& $168$\\
  \hline
  $\mu$ & $0$ & $0$ & $1$ & $1$ & $4$ & $6$ & $0$ & $8$ & $12$& $24$\\
  \hline
  $\nu$ & $0$ & $0$ & $0$ & $0$ & $2$ & $3$ & $1$ & $0$ & $0$ & $0$\\
  \hline
 \end{tabu}
\caption{Scalings of the homogeneous coordinates of the elliptically fibered CY threefold with base $\F{12}$.}
\label{tab:Scalings}
\tabulinesep=1.0mm
\end{table}
\begin{table}[t]
\tabulinesep=1.2mm
\centering
\begin{tabu}{|c|c|c||c||c|}
  \hline
  $\text{ord}(f)$ 	& $\text{ord}(g)$ 	& $\text{ord}(\Delta)$ 	& Name 				& \text{Gauge group}\\
  \hline
  0 			& 0 			& $n$ 			& $I_n$ 			& $\SU{n}$\\
  \hline
  $\geq 2$		& 3			& $n+6$			& \multirow{2}{*}{$I_n^{\,*}$}	& \multirow{2}{*}{$\SO{2n+8}$}\\
  \cline{1-3}
  2			& $\geq 3$		& $n+6$			& 				& \\
  \hline
  $\geq 3$ 		& 4 			& 8 			& $\fibiv^{\,*}$ 		& $\E6$\\
  \hline
  3 			& $\geq 5$ 		& 9 			& $\fibiii^{\,*}$ 		& $\E7$\\
  \hline
  $\geq 4$ 		& 5 			& 10 			& $\fibii^{\,*}$ 		& $\E8$\\
  \hline
 \end{tabu}
\caption{Excerpt from the Kodaira classification of the vanishing orders and the corresponding gauge groups.}
\label{tab:Kodaira}
\tabulinesep=1.0mm
\end{table}

\subsection{Heterotic -- F-theory duality in the smooth case}
\label{sec:DualitySmooth}
Before studying the singular limit we want to line out the duality in the smooth case which is much better understood \cite{Bershadsky:1996nh}. 

\subsubsection*{Fully Higgsed case with gauge group $\boldsymbol{\E8}$}
In the generic case the entire gauge group is Higgsed. Due to the fixed chirality (and the anomaly constraint \eqref{eq:6DGravAnomaly}) one loses one hypermultiplet per vector multiplet. Thus by assigning VEVs to the charged matter, i.e.\ to the $(\rep{56},\rep{1})$ and the $(\rep{1},\rep{2})$, we can break the visible sector gauge group completely, losing $133+3=136$ hypers and vectors. Since there is no charged matter under the second $\E8$, this group stays unbroken. Thus after Higgsing, we are left with $628-136=492$ singlets, 248 vector multiplets of the second $\E8$, and $1$ tensor multiplet. 

On the heterotic side, the 24 instantons are embedded in the whole $\E8$. Thus, we expect an $\E8$ bundle with instanton number 24 in the correspondence. Such a bundle has $472$ moduli. Together with the 20 geometric moduli of the K3, we recover the 492 singlets predicted from the F-theory side.

Let us see how to realize this situation in the Weierstrass model. In the generic case, $f$ and $g$ are arbitrary homogeneous polynomials in the coordinates $s,t,u,v$ of the base. We first expand them in $u,v$: 
\begin{subequations}
\label{eq:BasicExpansion}
\begin{align}
  \frac{1}{3}f &= c_{56} v^8 + c_{44} u v^7 + c_{32} u^2 v^6 + c_{20} u^3 v^5 + c_8 u^4 v^4\,,\\
  \frac{1}{2}g &= d_{84} v^{12} + d_{72} u v^{11} + d_{60} u^2 v^{10} + d_{48} u^3 v^9 + d_{36} u^4 v^8 +  d_{24} u^5 v^7 + d_{12} u^6 v^6 + d_0 u^7 v^5\,.
\end{align} 
\end{subequations}
There can be no higher terms because $u$ has a $\lambda$-weight of 12. The coefficients $c_i, d_j$ are homogeneous polynomials in $s,t$ where the degree is given by the subscripts; in particular, $d_0$ is just a number. With this general structure, we find for the discriminant  
\begin{align}
  \frac{1}{4\cdot27}\Delta&= f^3 +g^2= v^{10}\Delta_\text{red}\,.
\end{align}
Hence we have an $\E8$ singularity at $v=0$ and no further generic gauge group, since the reduced discriminant $\Delta_\text{red}$ does not factorize further and thus corresponds to an $I_1$ fiber degeneration, cf.\ table \ref{tab:Kodaira}.

In order to check the charged hypermultiplet spectrum, we investigate how the matter curves intersect. We see that the $I_1$ fiber locus $\Delta_\text{red}=0$ does not intersect the $\E8$ brane $v=0$, since
\begin{align}
  \Delta_\text{red}\sim \left(d_0^2 u^{14}+ \mathcal{O}\!\left(v\right)\right)\,.
\end{align}
Note that the monomial $u v$ is in the Stanley--Reissner ideal and thus the two coordinates cannot vanish simultaneously. Furthermore, $d_0$ is just a number not equal to zero: If it were zero, $\text{ord}(\Delta)=11$ at $v=0$, which is too singular to allow for a crepant resolution. So there are no ``brane intersections'', and we do not expect any charged matter. On the other hand, the neutral hypers can be counted by the coefficients in the the polynomials: The total number of coefficients in the $c_q$ and $d_k$ is 509. Subtracting three for $\text{SL}\!\left(2,\mathbbm{C}\right)$ acting on the base coordinates and 14 for $u\to\alpha u+\beta_{12}(s,t) v$, $v\to v/\alpha$, we end up with $509-3-14=492$. Hence, we reproduce the fully Higgsed heterotic model. 

Furthermore, we used \texttt{palp} \cite{Kreuzer:2002uu} as a cross-check to calculate the Hodge numbers of the elliptically fibered CY threefold over $\F{12}$. We get $h^{1,1}(X)=11$ and $h^{2,1}(X)=491$, and $h^{1,1}(\F{12})=2$. By comparing with \eqref{eq:FtheoryGeometrySpectrum}, we find indeed that the rank of the gauge group is $\text{rk}(V)=8$, and there are 492 neutral hypermultiplets.

\subsubsection*{Minimally Higgsed case with gauge group $\boldsymbol{\E7\times\E8}$}
The largest gauge group which can be obtained in perturbative string theory on the smooth heterotic side is when all instantons are embedded in the $\SU2$ (Alternatively, one can think of only giving VEVs to the $(\rep{1},\rep{2})$ but not to the $(\rep{56},\rep{1})$. This case corresponds to an \SU2 bundle which leaves an unbroken gauge group of $\E7\times\E8$. In the process of Higgsing, we break the $\SU2$ and lose three hypers. From the spectrum in table \ref{tab:OrbifoldSpectra}, we see that we end up with $2+8=10$ $\rep{56}$'s and $4+32\times2-3=65$ singlets in that case.

In order to get an enhanced gauge group from the Weierstrass model, we have to restrict the polynomials. It is rather straightforward to see that we can get an $\E7$ singularity if we set
\begin{align}
\label{eq:E7Weierstrass}
  c_{56}=c_{44}=c_{32}=0\,, \qquad d_{84}=d_{72}=d_{60}=d_{48}=d_{36}=0\,.
\end{align}
The determinant then factorizes as
\begin{align}
  \frac{1}{4\cdot27}\Delta&= u^9 v^{10}\Delta_\text{red}\,,
\end{align}
and we find an $\E8$ at $v=0$, an $\E7$ at $u=0$, and a smooth $I_1$ locus at $\Delta_\text{red}=0$ since it does not factorize further.

Regarding the matter spectrum, it is clear that on the one hand we have lost neutral hypermultiplets from the complex structure moduli (namely the ones in the polynomial we have set to zero), but on the other hand have gained charged matter from the resolution of the singularity enhancements where the ``branes'' intersect, i.e.\ where the singularity in the fiber gets worse. Explicitly, the remaining polynomials contain 69 parameters, but the reparameterizations that are left are just the ones on the base and a rescaling of $u$ (since $u=0$ should be fixed). Hence, we find $69-3-1=65$ neutral hypermultiplets. Regarding the charged matter, we look at the reduced determinant,
\begin{align}
\label{eq:deltared}
  \begin{split}
    \Delta_\text{red}	& =c_{20}^3 v^5 + \left(3 c_{20}^2 c_8 + d_{24}^2\right) u v^4 + \left(3 c_{20} c_8^2 + 2 d_{24} d_{12}\right)  u^2 v^3\\ 
			& \quad + \left(c_8^3 +  d_{12}^2 + 2 d_{24} d_0\right) u^3 v^2+ 2 d_{12} d_0 u^4 v + d_0^2 u^5\,, 
  \end{split}
\end{align}
which intersects $u=0$ at 20 points on the base $\P1$. At each of these intersection points, we expect one half-hypermultiplet of $\rep{56}$, yielding 10 full $\rep{56}$ hypers. Since again $u$ and $v$ cannot vanish simultaneously, we find no intersection of the $\E7$ with the $\E8$ curve or of the $\E8$ with the $I_1$ curve. 

From the cross-check with \texttt{palp} we obtain $h^{1,1}=18$ and $h^{2,1}=64$, which is consistent with our previous findings. We could now go back to the generic case by switching on VEVs, or polynomials, which is basically the path outlined in \cite{Bershadsky:1996nh}.

\subsection*{Completely unhiggsed case with gauge group $\boldsymbol{\E7\times\SU2\times\E8}$}
Motivated by the previous results, our ansatz is to factorize $\Delta_\text{red}$ further by choosing less generic polynomials such that we find an additional $I_2$ locus. In addition to the three extra vector multiplets, we should get three new hypermultiplets, which, together with the already-present 65 singlets of $\E7$ organize themselves into 32 doublets of $\SU2$ plus four remaining singlets. However, we expect the following problems: First, from the orbifold analysis we expect fractional instantons: we need a total instanton number of $24$, which should be divided evenly among the 16 $\Z2$ fixed points which are the only loci of non-zero curvature. Hence, each fixed point carries a fractional instanton number of $24/16=3/2$. Furthermore, these instantons, embedded in the first \E8, should not break the gauge group but only branch it to $\E7\times\SU2$. Second, the \SU2 doublets, the \E7 charged matter, and the instantons are localized at the orbifold fixed points. This leads to a singularity at the intersection points in codimension two which is too severe for the usual rules of calculating the spectrum to apply. Third, due to the homogeneity of the polynomials in $\Delta_\text{red}$, the extra \SU2 locus will be located at the zeros of a factor of the form $(u+p_{12} v)^2=0$, i.e.\ we get a natural quantization of matter states in multiples of 12 arising from the $\lambda$-scaling of $\F{12}$. However, the multiplicity of $(\rep{56},\rep{2})$ is 1 and the multiplicity of $(\rep{1},\rep{2})$ is 32, so neither is dividable by 12. Before we discuss how to tackle this problem in the next section, it is worthwhile to investigate other approaches to the duality to see how these problems arise there.

\subsection{Comparison with other singular K3 limits}
\label{sec:OtherConstructions}
In a series of beautiful papers \cite{Kumar:2009ac,Taylor:2010wm,Kumar:2010ru} a connection between the data from the anomaly polynomial and the geometry of the CY threefold was worked out. For the Green--Schwarz anomaly cancellation, the anomaly polynomial eight-form $\mathfrak{I}_8$ has to factorize into two four-forms as
\begin{align}
 \label{eq:6DAnomalyPolynomialFactorized}
 \mathfrak{I}_8=\left(\tr\, R^2-\sum_i\tr\, \alpha_i F_i^2\right)\left(\tr\, R^2-\sum_i\tr\, \widetilde{\alpha}_i F_i^2\right)\,,
\end{align}
where \tr is the trace in the fundamental and $R$ and $F$ are the curvature and the field strengths, respectively. In order to get to this factorization, the traces $\tr_{\rep{R}}$ in a representation $\rep{R}$ are written in terms of traces $\tr$ in the fundamental as
\begin{align}
\begin{split}
 \tr_{\rep{R}}\,F^2 &= A_{\rep{R}}\tr\,F^2\,,\\
 \tr_{\rep{R}}\,F^4 &= B_{\rep{R}}\tr\,F^4+C_{\rep{R}}\left(\tr\,F^2\right)^2\,.
\end{split}
\end{align}

\begin{table}[t]
 \centering
 \tabulinesep=0.8mm
 \begin{tabu}{|c|c||c|c|c|c|}
 \hline
 Gauge group			&Irrep $\rep{R}$&$A_\rep{R}$		&$B_\rep{R}$		&$C_\rep{R}$		& Normalization			\\
 \hline
 \multirow{2}{*}{\SU2}		&\rep{2}	&$1$			&$0$			&$\frac12$		& \multirow{2}{*}{$1$}		\\
 \cline{2-5}
				&\textbf{Adj}	&$4$			&$0$			&$8$			& 				\\
 \hline
 \multirow{2}{*}{\E7}		&\rep{56}	&$1$			&$0$			&$\frac{1}{24}$		& \multirow{2}{*}{$12$}		\\
 \cline{2-5}
				&\textbf{Adj}	&$3$			&$0$			&$\frac16$		& 				\\
 \hline
 \E8				&\textbf{Adj}	&$1$			&$0$			&$\frac{1}{100}$	& $60$				\\
 \hline
 \multirow{3}{*}{\SO{12}}	&\rep{12}	&$1$			&$1$			&$0$			& \multirow{3}{*}{$2$}		\\
 \cline{2-5}
				&\rep{32}	&$4$			&$-2$			&$\frac32$		& 				\\
 \cline{2-5}
				&\textbf{Adj}	&$10$			&$4$			&$3$			& 				\\
 \hline
 \end{tabu}
 \caption{Values of the $A_{\rep{R}}$, $B_{\rep{R}}$, and $C_{\rep{R}}$ used in the calculation of the anomaly polynomial.}
 \label{tab:GroupTheoryCoefficients}
\end{table}

The constants $A_{\rep{R}},B_{\rep{R}},C_{\rep{R}}$ can be calculated from group theory \cite{Erler:1993zy,Taylor:2010wm}, cf.\ table \ref{tab:GroupTheoryCoefficients}. Denoting the multiplicities of a state transforming in a representation $(\rep{R})$ by $x_{\rep{R}}$ and the multiplicities of a bi-fundamental transforming in a representation $(\rep{R},\rep{S})$ by $x_{\rep{R}\,\rep{S}}$, the real coefficients $\alpha,\widetilde{\alpha}$ in the factorized anomaly polynomial \eqref{eq:6DAnomalyPolynomialFactorized} have to fulfill
\begin{align}
\label{eq:alphaRelations}
\begin{split}
 \alpha_i+\widetilde{\alpha}_i&=\frac16\left(A_{\text{adj}}^i-\sum_{\rep{R}}x_{\rep{R}}^iA_{\rep{R}}^i\right)\,,\\
 \alpha_i\cdot\widetilde{\alpha}_i&=\frac23\left(\sum_{\rep{R}}x_{\rep{R}}^iC_{\rep{R}}^i-C_{\text{adj}}^i\right)\,,\\
 \alpha_i\cdot\widetilde{\alpha}_j+\alpha_j\cdot\widetilde{\alpha}_i&=4\sum_{\rep{R},\rep{S}}x_{\rep{R}\,\rep{S}}^{ij}A_{\rep{R}}^iA_{\rep{S}}^j\,.
\end{split}
\end{align}
Since the multiplicity of the states is related to the number of intersections of matter curves, the numbers $\alpha_i,\widetilde{\alpha}_i$ can be related to the divisors in $\F{m}$. Following the notation in table \ref{tab:Scalings}, we use the integral divisor basis $D_v=\{v=0\}$, $D_s=\{s=0\}\sim D_t=\{t=0\}$. Then $D_u=\{u=0\}=D_v+mD_s$ Their intersection numbers are
\begin{align}
\label{eq:FNIntersectionNumbers}
 D_v.D_v=-m\,,\qquad D_v.D_s=1\,,\qquad D_s.D_s=0\,.
\end{align}
and the anti-canonical divisor is
\begin{align}
 -K=D_u+D_v+D_s+D_t=2D_v+(2+m)D_s
\end{align}
The divisors $\xi_i$ corresponding to the matter curves are linked to $(\alpha_i,\widetilde{\alpha}_i)$ via
\begin{align}
 \xi_i=\frac{\alpha}{2}\left(D_v+\frac{m}{2}D_s\right)+\frac{\widetilde{\alpha}}{2}D_s\,.
\end{align}
Applying this to the ``minimally Higgsed'' case with gauge group $\E7\times\E8$ on $\F{12}$, one finds
\begin{align}
\begin{split}
\begin{aligned}
 (\alpha_{\E7},\widetilde{\alpha}_{\E7})&=(2,12) 	&&\Longrightarrow& \quad\xi_{\E7}&=D_v+12D_s=D_u\,,\\
 (\alpha_{\E8},\widetilde{\alpha}_{\E8})&=(2,-12)	&&\Longrightarrow& \quad\xi_{\E8}&=D_v\,.
\end{aligned}
\end{split}
\end{align}
Thus we find $\E7$ along $u=0$ and $\E8$ along $v=0$. For the matter spectrum one then finds using the intersection numbers \eqref{eq:FNIntersectionNumbers} that $\xi_{\E7}.\xi_{\E8}=0$. Furthermore, using that the discriminant  $\Delta=-12K$, we find for the reduced determinant $\Delta_{\text{red}}=-12\Delta-9\xi_{\E7}-10\xi_{\E8}=5D_v+60D_s$ where we used the vanishing orders given in table \ref{tab:Kodaira}. Thus $\Delta_{\text{red}}.\xi_{\E8}=0$ and $\Delta_{\text{red}}.\xi_{\E7}=60$, which corresponds to the $20$ zeros of $c_{20}^3$ in \eqref{eq:deltared} (counted with multiplicities).

Factorizing the orbifold anomaly polynomial $\mathfrak{I}_8$ for the case of $\E7\times\SU2\times\E8$ yields $(\alpha_{\SU2},\widetilde{\alpha}_{\SU2})=(\alpha_{\E7},\widetilde{\alpha}_{\E7})=(2,6)$ and thus $\xi_{\E7}=\xi_{\SU2}$. Consequently, one finds that $\xi_{\E7}.\xi_{\E8}=0=\xi_{\E8}.\xi_{\SU2}$ as it should be. Furthermore, we find $\xi_{\E7}.\xi_{\SU2}=12$, but there is only one state transforming in the bi-fundamental $(\rep{56},\rep{2})$. This factor of 12 is precisely the quantization in multiples of 12 noticed above. Furthermore, $\Delta_\text{red}=3D_v+36D_s$ and thus $\Delta_\text{red}.\xi_{\E8}=0$, which is as expected, but $\Delta_\text{red}.\xi_{\E7}=\Delta_\text{red}.\xi_{\SU2}=36$, while in the spectrum we have eight $(\rep{56},\rep{1})$ and 32 $(\rep{1},\rep{2})$.

From the orbifold point of view, this was bound to happen. So let us see why the multiplicity comes out incorrectly even though all anomalies cancel and the anomaly polynomial factorizes in the right way. The reason is that the normalization chosen in \eqref{eq:alphaRelations} is such that all instantons are integral. But as pointed out above, we have a fractional instanton number of $3/2$ at each fixed point. Without this change in normalization, one obtains
\begin{align}
\begin{split}
\begin{aligned}
 (\alpha_{\E7},\widetilde{\alpha}_{\E7})&=\left(\frac16,1\right)\,			&&\Longrightarrow\quad &\xi_{\E7} 	&=\frac{1}{12}D_v+D_s=\frac{1}{12}D_u\,,	\\
 (\alpha_{\SU2},\widetilde{\alpha}_{\SU2})&=\left(2,6\right) 				&&\Longrightarrow\quad &\xi_{\SU2}	&=D_v+12D_s=D_u\,,				\\
 (\alpha_{\E8},\widetilde{\alpha}_{\E8})&=\left(\frac{1}{30},-\frac{1}{5}\right)	&&\Longrightarrow\quad &\xi_{\E8}	&=\frac{1}{60}D_v\,.
\end{aligned}
\end{split}
\end{align}
It can be easily checked that these $\xi_i$ have the correct intersection numbers among each other, but they do not correspond to integral divisors in $\F{12}$.

However, we have not yet made use of the fact that the heterotic compactification space our duality construction is based on is singular itself and that the orbifold has discrete holonomy. In \cite{Aspinwall:1998xj}, this is accounted for by using the torsion part of the Mordell--Weil group. The authors give the form that $(f,g,\Delta)$ have to take such that the Mordell--Weil group contains a factor of $\Z{N}$ or $\Z{N}\times\Z{M}$. Then they consider the stable degeneration limit \cite{Friedman:1997yq,Aspinwall:1997ye}. We will not go into the details of this construction and simply mention that in the stable degeneration the Hirzebruch surface $\F{m}$ degenerates into two Hirzebruch surfaces $\mathbbm{F}^i_{m}$ which intersect along a curve $C_*$. This curve forms together with its fibers the heterotic K3 in the large volume limit. Cast into the language of divisors (or their dual curves, which is used interchangeably by abuse of notation), this means we have divisors $C_0$ and $C_*$ in $\mathbbm{F}^1_{m}$ and divisors $C_\infty$ and $C_*$ in $\mathbbm{F}^2_{m}$, where the two $C_*$'s have to be identified. The $\Z{N}$ singularities of the heterotic K3 surface arise from intersections of $I_N$ curves with $C_*$. We are interested in the case where the Mordell--Weil group contains a $\Z2$ factor, which has also been analyzed in \cite{Marquart:2002bz}. We place the $\E7$ along $\xi_{\E7}=C_0$. Using the form of the discriminant necessary to get a $\Z2$ factor that gives rise to the $(\E7\times\SU2)/\Z2$ gauge group \cite{Aspinwall:1998xj}, one finds that the $I_2$ curve is along $\xi_{\SU2}=D_v+8D_s$. This intersects $C_*\sim D_u$ eight times (no matter what $m$ is). Hence this construction leads to the ``wrong'' singular K3 limit (i.e.\ not the one dual to the orbifold which has 16 $\Z2$ singularities). By looking at the Euler number, we can see what is happening: the limit corresponds to K3$/\Z2$ rather than $T^4/\Z2$. To show this let us denote the number of fixed points by $k$. We can calculate the Euler number of the (smooth) K3 \cite{GSW2} by starting from the original singular space $X$, subtracting the number of fixed points, dividing by the $\Z2$ action, and gluing back in $k$ exceptional divisors $E$:
\begin{align}
 \chi(\text{K3}^{\text{smooth}})=\frac{\chi(X)-k}{|\Z2|}+k\chi(E)\,.
\end{align}
Using $\chi(\text{K3}^{\text{smooth}})=24$, $\chi(E)=2$, $\chi(X=\text{K3})=24$, $\chi(X=T^4)=0$, we find that $k=8$ for $X=\text{K3}$ and $k=16$ for $X=T^4$. This means that the global compactification obtained by using this method does not result in the heterotic dual we want to construct\footnote{The same happens for the other orbifolds. In the $\Z3$ case for example, one can show that this method produces only $6$ fixed points instead of the $9$ that we would expect from the orbifold, which is again consistent with a quotient K3$/\Z3$ rather than $T^4/\Z3$.}.


\section{F-theory duals of heterotic orbifold models}
\label{sec:OrbifoldDuals}

After having outlined why the previously considered constructions cannot work in our case we now describe the approach we are following instead. In order to do so it is thus instructive to take a closer look at the duality \cite{Morrison:1996na,Morrison:1996pp}. As explained in figure \ref{fig:CYFibration}, we have a double fibration structure on the F-theory side. In particular, the heterotic K3 is essentially the base $\P1$ together with the elliptic fiber. This means that the ``middle'' terms in $f$ and $g$, i.e.\ the terms $c_8 u^4 v^4$ and $d_{12} u^6v^6$ should contain the geometric moduli of the heterotic K3. Indeed, subtracting an overall scaling, we obtain $8+12=20$ moduli. Likewise, the zeros of $d_{24}$ and $d_0$ correspond to the instantons in the two $\E8$'s. Again, after subtracting an overall scaling, one obtains $24$ and $0$, as it should be. To describe the fibration structure depicted in figure \ref{fig:CYFibration}, we use similar methods to the ones described in \cite{Braun:2009wh}.

\subsection[Constructing the duals of the \texorpdfstring{$T^4/\Z2$}{T4/Z2} orbifold]{Constructing the duals of the \texorpdfstring{$\boldsymbol{T^4/\Z2}$}{T4/Z2} orbifold}
\label{sec:T4Z2DualConstruction}
Pictorially, the $T^4/\Z2$ orbifold is a pillow with a torus over it everywhere except at the corners, where the fiber is a pillow itself, cf.\ figure \subref*{subfig:T4Z2Alternative}. Any section will hit one of the fiber pillow corners at the base pillow corners. On the other hand, a Weierstrass fiber torus always has one singularity (at $\Delta=0$), not four, and a non-singular section at $z=0$. Hence we have to deform the orbifold in order to write it as a Weierstrass model. The method of \cite{Braun:2009wh} is to
\begin{itemize}
  \item pick one section,
  \item blow up the fiber singularities it hits (such that we basically have a pillow where one of the corners is blown up to a smooth $\P1$),
  \item blow down the other finite-volume component of the fiber (i.e.\ the original pillow).
\end{itemize}
\begin{figure}[t]
 \centering
 \includegraphics[width=0.9\textwidth]{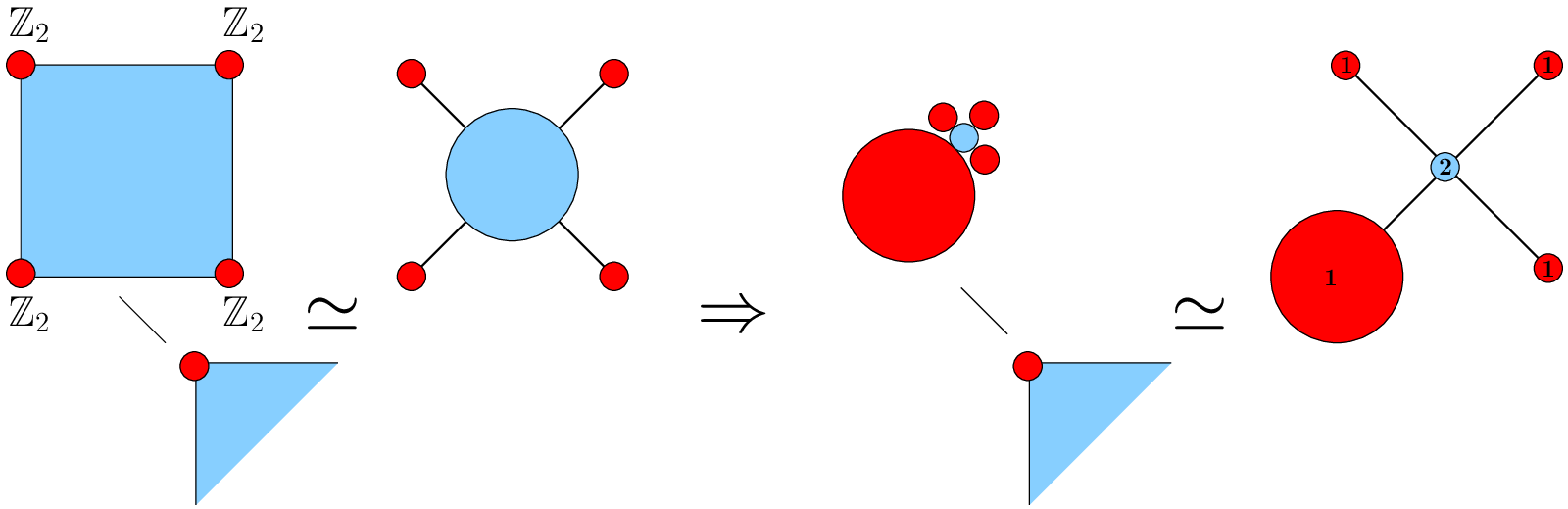}
 \caption{Blowdown limit of $T^4/\Z2$. Over each base singularity there is an orbifold pillow with four $\Z2$ singularities, corresponding to the affine Dynkin diagram of $\SO8$. In the blowdown procedure, the central torus node (blue) is blown down and one of the corner singularities (red) is blown up.}
 \label{fig:T4Z2BlowdownLimit}
\end{figure}
This procedure is summarized pictorially in figure \ref{fig:T4Z2BlowdownLimit}. In this way we end up with a fiber with a $D_4$ singularity (made up of three of the original pillow corners and the blown-down pillow itself), where the section sits at a smooth point. In particular, the affine node corresponds to one of the four corner singularities and the original orbifold pillow corresponds to the central (adjoint) node. Note that the central node has multiplicity (or Dynkin label) two. Hence it is intersected by a two-section rather than a section. In contrast, after applying the above procedure, we obtain a finite component with multiplicity one, which can hence be hit by a section and thus should have a Weierstrass model. This means we need to find a complex two-dimensional Weierstrass model with four $D_4$ singularities. We start from the heterotic Weierstrass model
\begin{align}
 \label{eq:HeteroticWeierstrassEq}
  y^2&=x^3 + \widetilde{f}_{8} x z^4 + \widetilde{g}_{12} z^6\,,
\end{align}
where we identify $\widetilde{f}_{8}=c_{8}$ and $\widetilde{g}_{8}=d_{12}$ and require the vanishing orders of $(\widetilde{f},\widetilde{g},\widetilde{\Delta})$ to be $(2,3,6)$ at four points in the base (cf.\ table~\ref{tab:Kodaira}). This fixes the Weierstrass equation to
\begin{align}
  y^2&=x^3 + \widetilde{\alpha} p_4^2 x z^4 + \widetilde{\beta} p_4^3 z^6\,.
\end{align}
Here $p_4$ is a polynomial of degree four in $s,t$, whose four zeros correspond to the position of the four fixed points in the base. The discriminant is
\begin{align}
  \frac{1}{4\cdot 27}\,\widetilde{\Delta}=\left(\widetilde{\alpha}^3+\widetilde{\beta}^2\right) p_4^6\,,
\end{align}
so we clearly have four $D_4$ singularities at the roots of $p_4$. As an additional check, note that the complex structure $\tau$ is given in terms of the $j$-function by
\begin{align}
  \label{eq:CSHeteroticTorus}
  j\!\left(\tau\right)\sim \frac{\widetilde{f}^3}{\widetilde{\Delta}}\sim \frac{\widetilde{\alpha}^3}{\left(\widetilde{\alpha}^3+\widetilde{\beta}^2\right)}\,,
\end{align}
i.e.\ it is constant but not fixed. This fits nicely with the $\Z2$ orbifold, in which the complex structure is also not fixed since every lattice has a reflection symmetry.

This orbifold argument fixes $c_8=\widetilde{\alpha}p_4^2$ and $d_{12}=\widetilde{\beta}p_4^3$. Furthermore, we know that the $24$ instantons are located at the fixed points as well, which suggests $d_{24}=\widetilde{\gamma}p_4^6$. Looking at \eqref{eq:deltared}, we see that we have fixed all polynomials except for $c_{20}$. Requiring that we find an extra $\SU2$ symmetry forces us to take $c_{20}=\widetilde{\kappa} p_4^5$ and to relate the numerical coefficients $\widetilde{\alpha},\widetilde{\beta},\widetilde{\gamma},\widetilde{\kappa}$ among each other. By rescaling and leaving $d_0$ and the coefficients in $p_4$ as free parameters, we then find
\begin{subequations}
  \begin{align}
    \frac{1}{3}f &=\frac{4}{d_0} u^3 v^4 p_4^2\left(d_0 u+ 12 p_4^3 v\right)\,,\\
    \frac{1}{2}g &=\frac{1}{d_0}  u^5 v^5 \left(d_0^2 u^2-4d_0p_4^3u v-108p_4^3 v^2\right)\,,\\
    \frac{1}{4\cdot27} \Delta &= \frac{1}{d_0^3} u^9 v^{10} \left(d_0u + 6 p_4^3 v\right)^2 \left(d_0^3 u^3+20 d_0^2p_4^3 u^2v+68d_0p_4^6 u v^2+3072p_4^9v^3\right)\,.
  \end{align}
\end{subequations}
Clearly, there is an extra locus of type $I_2$ fibers over $d_0u+6p_4^3 v=0$. 

\subsubsection*{Spectrum}
When looking at the spectrum, one has to be very careful as results may be different from what one naively expects. We present a discussion of the subtleties from different point of views. First we find that (by construction) still nothing intersects the $\E8$ curve at $v=0$. However, the price to pay is that the $\E7$ curve, the $\SU2$ curve, and the $I_1$ curve all intersect at the four points $u=0=p_4$. This is to be expected since all matter (except for the untwisted $(\rep{56},\rep{2})$ and the four singlets) is localized at the orbifold fixed points located at the zeros of $p_4$ in the base.

Now comes the tricky part: the blowup of the singularities in the base seems to introduce new tensor multiplets (whose singlets correspond to the blowup moduli), which would result in models with $N_\text{T}\neq1$ and thus would not be dual to a perturbative heterotic model. However, as argued by Aspinwall and Donagi in \cite{Aspinwall:1998he}, the spectrum computation is much more subtle in this case. Naively one might have expected that the orbifold is the limit in which the gauge bundle becomes concentrated at the orbifold singularities, leading to point-like instantons. However, these point-like instantons are not the same as the tangent sheaf (albeit having the same support). The difference becomes apparent when looking at the spectral curve. In the former case it contains a reduced component while in the latter case it contains a fat line. This leads to different effects when resolving the singularities. In particular, in order to decide which of the multiplets (tensor, gauge, and matter multiplets) that are expected from the resolution actually do occur, one needs to investigate the corresponding extremal transitions. By studying a similar case, the authors of \cite{Aspinwall:1998he} find that the occurrence of extra tensor multiplets is indeed blocked, while the \SU2 together with the four half-hyper doublets do occur.

The fact that the spectrum differs from the naive expectations can also be understood from the heterotic/M-Theory duality. Note that the instantons at the orbifold fixed points cannot be ``ordinary'' point-like instantons which would correspond to M5 branes. These ordinary instantons do not break or branch the gauge group; the corresponding new tensor multiplets balance the vector multiplets in the anomaly \eqref{eq:6DGravAnomaly}. This fits well with the fact that point-like instantons usually have trivial holonomy \cite{Aspinwall:1996mn,Aspinwall:1998he}. In contrast, point-like instantons at orbifold singularities are expected to inherit the orbifold holonomy, which is fractional. This means that these instantons can (and do) branch the gauge group, removing some $W$-bosons from the spectrum. In order to still satisfy the anomaly constraint \eqref{eq:6DGravAnomaly}, no new tensor multiplets do occur. Hence these point-like instantons with non-trivial fractional holonomy cannot correspond to ordinary M5 branes. In some sense, these M5 branes with fractional holonomy are forced to sit at the fixed point with the same fractional holonomy and cannot travel though the bulk (which has trivial holonomy); hence they cannot be removed from the singularity and become ordinary M5 branes. The presence of spaces with singularities were studied in an M-theory description in \cite{Witten:1997bs,deBoer:2001px}. There, the singularities correspond to ``frozen singularities'' with discrete 3-form flux obstructing their resolution. It would be very interesting to construct these M-theory models directly and compare with the Jacobian of the F-theory models as discussed in \cite{Braun:2014oya}, which is however beyond the scope of the paper.

Despite these complications, we demonstrate that the spectrum can be determined based on counting arguments as it was done in section \ref{sec:DualitySmooth} and is found to be as expected from the perturbative heterotic orbifold point of view. To determine the number of $\rep{56}$'s, we note that after deforming the extra $\SU2$ locus away by switching on coefficients in the polynomial, we are again in the ``minimally Higgsed'' case described in section \ref{sec:DualitySmooth} with $20$ $\rep{56}$'s, which assemble into sixteen half hypers $(\rep{56},\rep{1})$ and one full hyper $(\rep{56},\rep{2})$. The uncharged moduli correspond to the parameters left in the polynomials. On the heterotic side, there are four (two are related to the size and two are related to the complex structure of the $T^4=T^2\times T^2$ underlying the $T^4/\Z2$ orbifold. On the F-theory side we find that there are $9$ parameters: $\widetilde{\alpha}$, $\widetilde{\beta}$, $\widetilde{\gamma}$, $\widetilde{\kappa}$, $d_0$ and four parameters of $p_4$. However, the zeros of $p_4$ are fixed as they correspond to the orbifold fixed point locations. In addition, the other parameters have to be related amongst each other for the extra $\SU2$ locus to appear, which leaves us with four free parameters. In particular, the complex structure of the base torus depends on a combination of $\widetilde{\alpha}$ and $\widetilde{\beta}$, cf.\ \eqref{eq:CSHeteroticTorus}. Finally, the doublets of $\SU2$ are related to those deformations of the original polynomials $c_{20}$, $c_{8}$, $d_{24}$, $d_{12}$, and $d_{0}$ that destroy the $\SU2$ locus. As discussed in section \ref{sec:DualitySmooth}, these have $69$ parameters out of which one is an overall scaling, so there are $68$. Out of these four respect the $\SU2$ symmetry (i.e.\ the four singlets), so $64$ parameters destroy the extra $\SU2$ locus. These $64$ correspond to the 64 half-hypers $(\rep{1},\rep{2})$.

\subsubsection*{Stable degeneration}
In terms of the stable degeneration limit, this means that the 24 $I_1$ fibers that generically intersect the divisor $C_*$ are grouped into four groups with 6 $I_1$'s each, leading to four $I_0^*$ singularities. Since the discriminant vanishes with order 1 along an $I_1$ and with order $6$ along an $I_0^*$, we see that all 24 $I_1$ fiber components are used in this limit. Along the divisors $C_0$ we keep an $\E8$ fiber and along the divisor $C_\infty$ we split the $\E8$ into $\E7\times\SU2$. When blowing up the central node in each of the $I_0^*$, we obtain the 16 $I_2$ singularities of the $T^4/\Z2$ orbifold. It is not possible to obtain 16 $I_2$ directly, as this would require $32$ $I_1$ fibers, but in a K3 there are only 24.

Of course the techniques outlined here can also be applied to other $T^4/\Z{N}$ orbifolds, which we will discuss now in turn.

\subsection[Constructing the duals of the \texorpdfstring{$T^4/\Z3$}{T4/Z3} orbifold]{Constructing the duals of the \texorpdfstring{$\boldsymbol{T^4/\Z3}$}{T4/Z3} orbifold}
\label{sec:T4Z3DualConstruction}

\begin{figure}[t]
 \centering
 \includegraphics[width=0.45\textwidth]{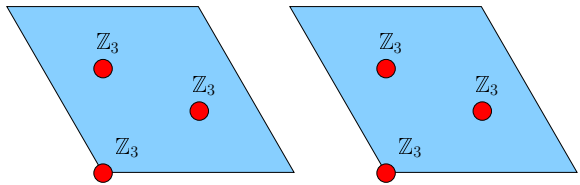}
 \caption{The orbifold $T^4/\Z3$ has three $\Z3$ singularities in each $T^2$.}
 \label{fig:T4Z3Orbifold}
\end{figure}

We start with discussing the orbifold. It has nine $\Z3$ (or $A_2$) fixed points (three per $T^2$ plane, cf.\ figure \ref{fig:T4Z3Orbifold}) and two twisted sectors, 
\begin{align}
\theta\,:~~(z_1,z_2)\mapsto(e^{2\pi\i/3}z_1,e^{-2\pi\i/3}z_2)\,,\qquad \theta^2\,:~~(z_1,z_2)\mapsto(e^{-2\pi\i/3}z_1,e^{2\pi\i/3}z_2)\,.
\end{align}
This means that if we distribute the instantons again evenly, we have $n_3=24/9=8/3$ instantons per fixed point or $4/3$ per resolution $\P1$. In order to be able to mod out the $\Z3$ orbifold action, the complex structures of the two $T^2$ have to be fixed to $\tau=e^{2\pi\i/3}$. For the spectrum, this means that there are two moduli (completely uncharged singlets) corresponding to the overall sizes of the tori. Since the commutant of $\E8$ with $\Z3$ is $\U1$, we find a gauge group of $\E7\times\U1\times\E8$. Each fixed point has one $\rep{56}$ and seven singlets, both charged under the $\U1$. Furthermore, there is another $\rep{56}$ and another charged singlet in the untwisted sector. It can easily be checked that this satisfies the anomaly constraint \eqref{eq:6DGravAnomaly} (as well as all other anomaly constraints, of course).

\begin{figure}[t]
 \centering
 \includegraphics[width=0.9\textwidth]{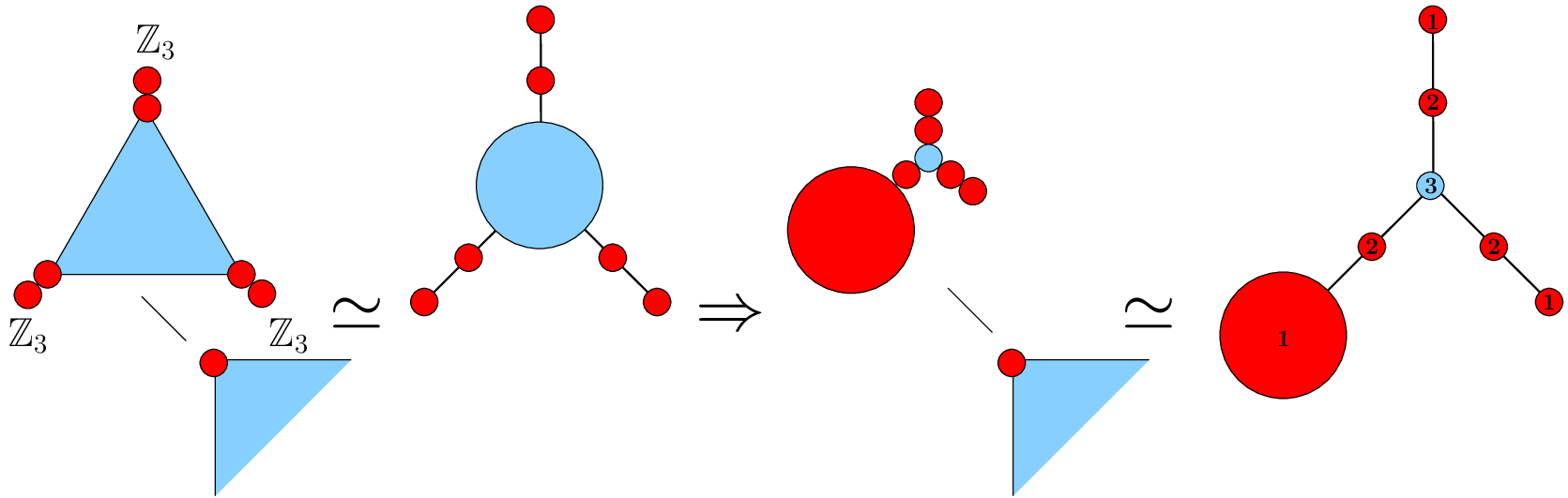}
 \caption{Blowdown limit of $T^4/\Z3$. Over each base singularity there is an orbifold pillow with three $\Z3$ singularities, corresponding to the affine Dynkin diagram of $\E6$. In the blowdown procedure, the central torus node (blue) is blown down and one of the corner singularities (red) is blown up.}
 \label{fig:T4Z3BlowdownLimit}
\end{figure}
In the case where the $\U1$ is broken, the spectrum is the same as for the $\Z2$ model with broken \SU2 (ten \rep{56}'s and $9\cdot7+2=65$ singlets), so we can take the same polynomials at that stage. By the same argument as above, we cannot directly take the orbifold as a Weierstrass model, but have to perform a similar blowup and blowdown.  This time, however, the procedure takes us from nine $A_2$ singularities to three $\E6$ ones, cf.\ figure \ref{fig:T4Z3BlowdownLimit}. The Dynkin label of the original finite (central) component is $3$ and thus should be intersected by a three-section, while after applying the procedure, we obtain a finite component with multiplicity 1, which can be described as a Weierstrass model with a single section. The vanishing orders of $\widetilde{f}_8$ and $\widetilde{g}_{12}$ in the heterotic Weierstrass equation \eqref{eq:HeteroticWeierstrassEq} should be $\geq3$ and $4$ at three points to obtain an \E6. But clearly $\widetilde{f}_8$ is not the cube of any degree-three polynomial, so we have to set $\widetilde{f}_8=0$ (note that indeed the vanishing order for $\E6$ is $\text{ord}(f)\geq 3$, not necessarily equal) On the other hand, we can easily set $\widetilde{g}_{12}=\alpha p_3^4$ to obtain a vanishing order of 4 at three points in the base. This has the nice effect that the the complex structure of the fiber is given by
\begin{align}
  j\!\left(\tau\right)\sim \frac{\widetilde{f}_8}{\widetilde{\Delta}}=0\,,\widetilde{f}_8
\end{align}
i.e.\ it is fixed to $\tau=e^{2\pi\i/3}$, just as it should be for the $\Z3$ orbifold.

Hence we have fixed $c_8$ and $d_{12}$. Going further, we note the interpretation of $d_{24}$ as the instanton location implies that we should have $d_{24}\sim p_3^8$, since we have 24 instantons equally divided between the three singularities, i.e.\ eight per singularity. On the other hand, $c_{20}$ is again not fixed. However, by comparing to the minimally Higgsed case and to the $\Z2$ orbifold results, we find that each zero of $c_{20}$ contributes one $\rep{56}$ half-hypermultiplet, which have to combine into full hypermultiplets in this case since their \U1 charge forces them into complex representations, so we make the ansatz $c_{20}\sim p_3^6 k_2$. The zeros of $k_2$ should then correspond to the untwisted $\rep{56}$, while the $p_3^6$ gives rise to the twisted $\rep{56}$'s. As in the $\Z2$ case, we find that there are $18$ half-hypers of $\rep{56}$, one per $\P1$ needed to resolve the $\Z3$ singularity. Finally, the factorization should be such that there is an extra section leading to the \U1 symmetry, but a detailed exploration of this is beyond the scope of this paper. The parameter counting is similar to the $\Z2$ case (since they both have the same minimally Higgsed limit), except that there are two parameters less (corresponding to the complex structures of the two $T^2$, which are fixed for this orbifold).

\subsubsection*{Stable degeneration}
In this case the 24 $I_1$ fibers are grouped into three groups with 8 $I_1$'s each, resulting in three $\fibiv^*$ singularities intersecting $C_*$. This uses up again all 24 $I_1$ fibers. Along the divisors $C_0$ we keep an $\E8$ fiber and along the divisor $C_\infty$ we split the $\E8$ into $\E7\times\U1$. When blowing up the central node in each $\fibiii^*$, we obtain the 9 $I_3$ singularities of $T^4/\Z3$. Again, obtaining 9 $I_3$ singularities directly is impossible, since it would require 27 $I_1$ fibers.

\subsection[Constructing the duals of the \texorpdfstring{$T^4/\Z4$}{T4/Z4} orbifold]{Constructing the duals of the \texorpdfstring{$\boldsymbol{T^4/\Z4}$}{T4/Z4} orbifold}
\label{sec:T4Z4DualConstruction}
\begin{figure}[t]
 \centering
 \includegraphics[width=0.4\textwidth]{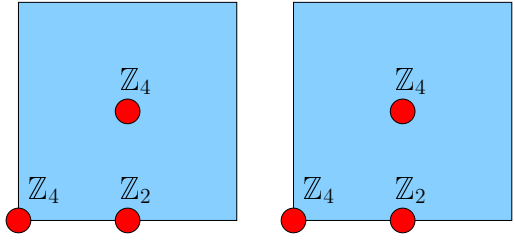}
 \caption{The orbifold $T^4/\Z4$ has two $\Z4$ singularities and one $\Z2$ singularity in each $T^2$.}
 \label{fig:T4Z4Orbifold}
\end{figure}

\begin{figure}[t]
 \centering
 \includegraphics[width=0.9\textwidth]{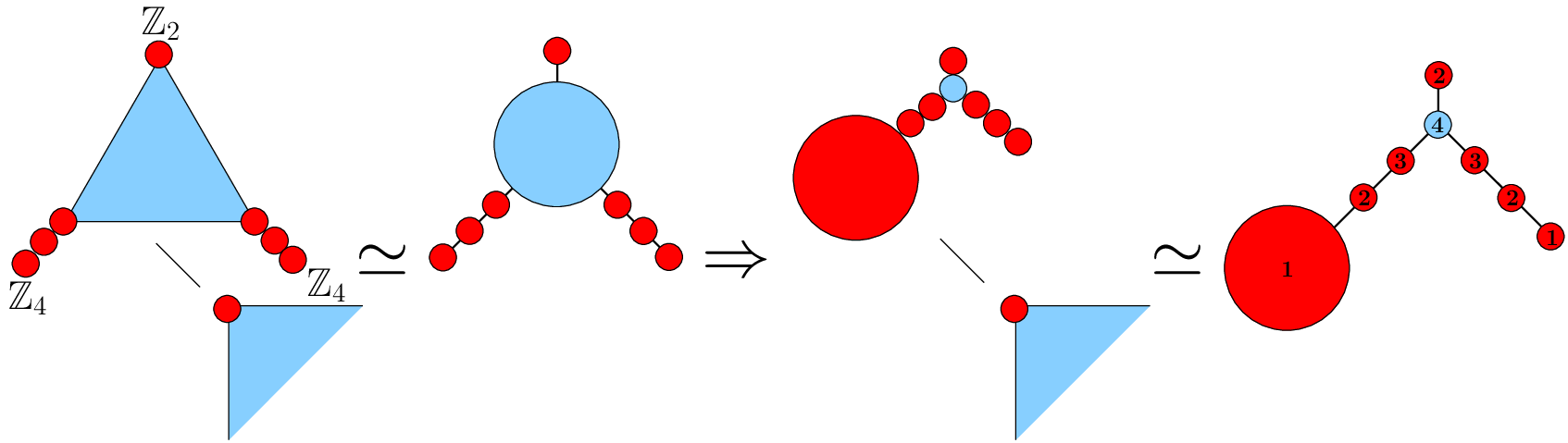}
 \caption{Blowdown limit of $T^4/\Z4$. Over a base singularity there is an orbifold pillow with two $\Z4$ (one of which is depicted here) and one $\Z2$ (depicted in figure \ref{fig:T4Z2BlowdownLimit})  singularities, corresponding to the affine Dynkin diagram of $\E7$. In the blowdown procedure, the central torus node (blue) is blown down and one of the corner singularities (red) is blown up.}
 \label{fig:T4Z4BlowdownLimit}
\end{figure}

This orbifold has three twisted sectors,
\begin{align}
\theta:~(z_1,z_2)&\mapsto(e^{2\pi\i/4}z_1,e^{-2\pi\i/4}z_2)\,,~~ \theta^2:~(z_1,z_2)\mapsto(-z_1,-z_2)\,,~~\theta^3=\theta^{-1}
\end{align}
but the third sector is the CPT conjugate of the first, so it is enough to discuss the first and second twisted sector. In the first sector, we have two $\Z4$ fixed points (which are at the same time $\Z2$ fixed points of the second twisted sector) and one $\Z2$ fixed point (cf.\ figure \ref{fig:T4Z4Orbifold}). For the underlying $T^2$'s to be compatible with the $\Z4$ action, the complex structure of the tori have to be fixed to $\tau=\i$. Picking one $T^2$ as the base and the other one as the fiber as before, this means that there are 
\begin{itemize}
  \item two points on the base over which there is a ``$\Z4$ fiber'', with two $\Z4$ and one $\Z2$ singularities, which in blowup lead to an extended $\E7$ diagram (cf. figure \ref{fig:T4Z4BlowdownLimit}),
  \item and one point in the base over which we have the known ``$\Z2$ fiber'', leading to the extended $D_4$ diagram (cf.\ figure \ref{fig:T4Z2BlowdownLimit}).
\end{itemize}
Again, this takes us from a finite component with multiplicity $4$ to a finite component with multiplicity $1$ in the $\Z4$ fibers; for the $\Z2$ fiber, the story is as in the $T^4/Z_2$ case. The required vanishing orders of $(\widetilde{f}_8,\widetilde{g}_{12},\widetilde{\Delta})$ for the Weierstrass model on the heterotic side are $(3,\geq5,9)$ for $\E7$ and $(2,\geq3,6)$ for $D_4$. Since there are two $\Z4$ and one $\Z2$ fiber, the combined varnishing orders are $(8,13,24)$. However, $\widetilde{g}_{12}$ is a section in $12 D_u$; hence it cannot vanish to order 13 and we have to set $\widetilde{g}_{12}=0$ identically. This means that the complex structure of the fibration is again fixed since $\widetilde{\Delta}=4\widetilde{f}^3$, and thus
\begin{align}
  j\!\left(\tau\right)\sim \frac{\widetilde{f}^3}{\widetilde{f}^3}= 1
\end{align}
From the properties of the $j$~function, this implies that indeed $\tau=\i$. 

In summary, we have
\begin{align}
  c_8=\widetilde{f}_8= \alpha p_1^3 q_1^3 r_1^2\,, \qquad d_{12}=\widetilde{g}_{12}=0\,.
\end{align}
Here, $p_1,q_1$, and $r_1$ are all polynomials of degree one, giving the locations of the two $\E7$ and the $D_4$ fiber, respectively. Again, the instantons are located at the singular fibers, so
\begin{align}
  d_{24}= \beta p_1^i q_1^i r_1^j
\end{align}
with $2i+j=24$. Since $r_1=0$ is a $\Z2$ fiber, we may assume just as for the $\Z2$ orbifold above that $j=6$ and thus $i=9$, i.e.\ nine instantons per $\Z4$ fiber. Using that the $\Z2$ fixed points carry $n_2=\frac{3}{2}$ instantons, we find that each $\Z4$ fixed point carries $n_4=\frac{15}{4}$ instantons (and thus $\frac54$ instantons per resolution $\P1$), such that $2n_4+n_2=9$. We thus find that $\Z4$ fixed points carry fractional (quarter) instantons, as seems reasonable. 

Finally, using again that the half-hypers $\rep{56}$ are localized at the zeros of $c_{20}$, we expand it in powers of $p_1,q_1,r_1$. Assuming that each $\P1$ needed in the resolution process comes with one twisted $\rep{56}$ half-hyper as in the previous cases, we find that 
\begin{align}
  c_{20}=\gamma p_1^7 q_1^7 r_1^4 k_2\,,
\end{align}
where again $k_2$ is a degree-two polynomial accounting for the untwisted $\boldsymbol{56}$, which cannot be composed out of two half-hypers because of its $\U1$ charge. Note, however, that $p_1$ and $q_1$ are raised to odd powers. This means that some of these have to correspond to actual half-hypermultiplets, which consequently cannot be charged under \U1. The contributions to the massless spectrum are as follows:
\begin{itemize}
  \item A $\Z4$ fixed point contains three $\P1$'s, one of which can be though of as a $\P1$ corresponding to the $\Z2$ fixed point (i.e.\ belonging to the second twisted sector). Thus, each $\Z4$ point contains a full $\boldsymbol{56}$ for the first twisted sector and a half $\boldsymbol{56}$ for the $\theta^2$ sector.
  \item The pure $\Z2$ fixed points simply contain half $\boldsymbol{56}$'s, corresponding to the $\theta^2$ sector.
\end{itemize}
Adding these up, we find four full $\rep{56}$'s in the $\theta$ sector and $4+2(1+1+1)=10$ half-hypers in the $\theta^2$ sector, which matches precisely the spectrum of the orbifold, cf.\ table \ref{tab:OrbifoldSpectra} and the discussion of the distribution of the $\rep{56}$'s in section \ref{sec:6DHeteroticModels}. The counting of parameters is as in the $\Z3$ case. Again, we find that there are only two uncharged singlets (geometric moduli), corresponding to the sizes of the tori; the other two singlets corresponding to the complex structure are lost since it had to be fixed to be compatible with the $\Z4$ action.

\subsubsection*{Stable degeneration}
Here  we group the 24 $I_1$ fibers into two groups with 9 $I_1$'s each and one group with 6 $I_1$'s, resulting in two $\fibiii^*$ and one $I_0^*$ singularities intersecting $C_*$. This uses up again all 24 $I_1$ fibers. As before, we keep the $\E8$ fiber along $C_0$ and split the other $\E8$ along $C_\infty$ into $\E7\times\U1$. When blowing up the central nodes, we obtain 4 $I_4$ and 6 $I_2$ singularities. Again, obtaining these singularities directly would require 28 $I_1$ fibers intersecting $C_*$.

\subsection[Constructing the duals of the \texorpdfstring{$T^4/\Z6$}{T4/Z6} orbifold]{Constructing the duals of the \texorpdfstring{$\boldsymbol{T^4/\Z6}$}{T4/Z6} orbifold}
\label{sec:T4Z6DualConstruction}
\begin{figure}[t]
 \centering
 \includegraphics[width=0.45\textwidth]{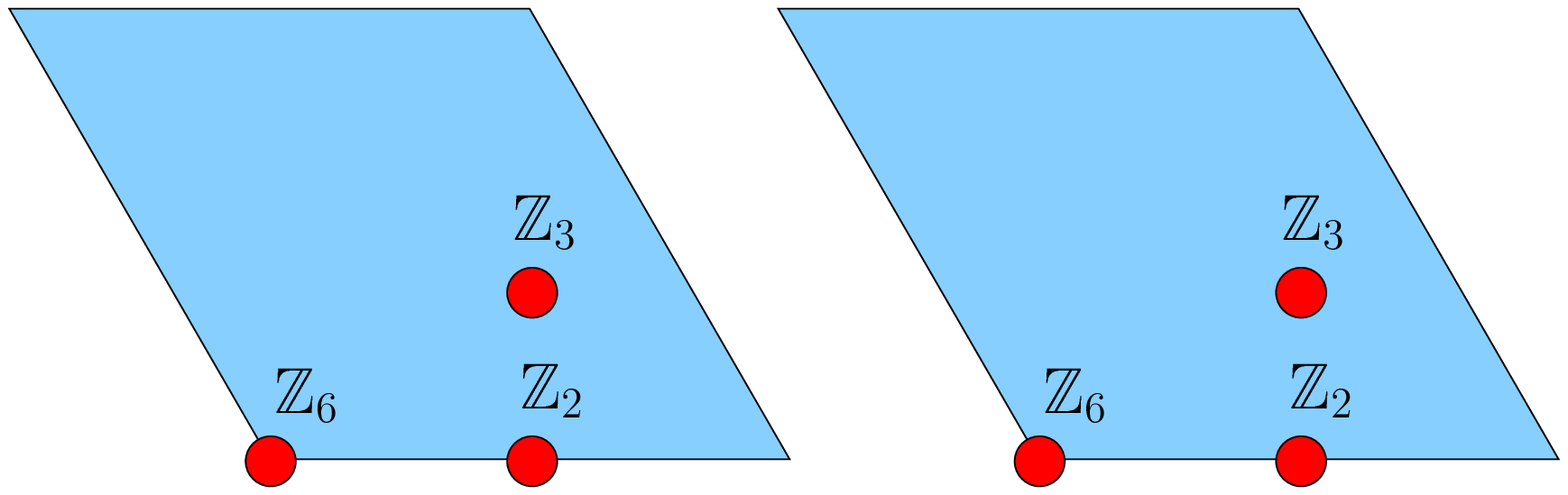}
 \caption{The orbifold $T^4/\Z6$ has one $\Z6$, one $\Z3$, and one $\Z2$ singularity in each $T^2$.}
 \label{fig:T4Z6Orbifold}
\end{figure}

Lastly, we want to discuss the $\Z6$ orbifold. Here, the orbifold action introduces one fixed point of each order $\Z6$, $\Z3$ and $\Z2$ (cf.\ figure \ref{fig:T4Z6Orbifold}) in each torus,
\begin{align}
\begin{split}
\theta\,:~~(z_1,z_2)&\mapsto(e^{2\pi\i/6}z_1,e^{-2\pi\i/6}z_2)\,,\qquad \theta^2\,:~~(z_1,z_2)\mapsto(e^{2\pi\i/3}z_1,e^{-2\pi\i/3}z_2)\,,\\ 
\theta^3\,:~~(z_1,z_2)&\mapsto(z_1,-z_2)\,,\qquad\theta^4=\theta^{-2}\,,\qquad\theta^5=\theta^{-1}\,.
\end{split}
\end{align}
The $\theta^4$ sector is the CPT conjugate of the $\theta^2$ sector, and the $\theta^5$ sector is the CPT conjugate of the $\theta$ sector, so it is enough to discuss only the first three twisted sectors.

\begin{figure}[t]
 \centering
 \includegraphics[width=0.9\textwidth]{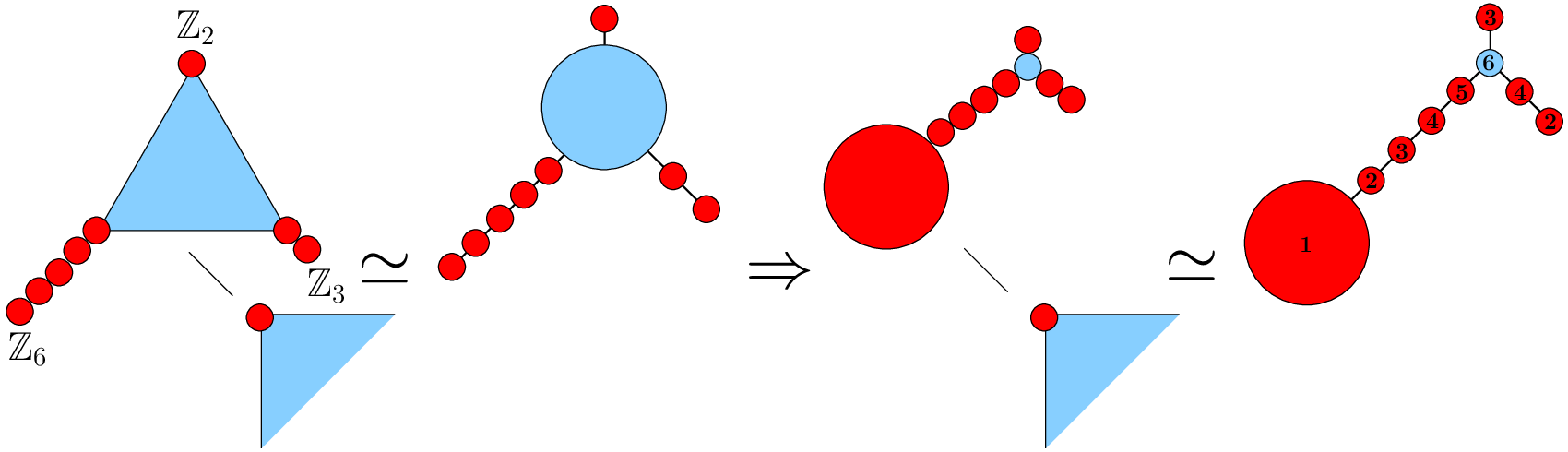}
 \caption{Blowdown limit of $T^4/\Z6$. Over a base singularity there is an orbifold pillow with one $\Z6$ (depicted here), one $\Z3$ (depicted in figure \ref{fig:T4Z3BlowdownLimit}), and one $\Z2$ (depicted in figure \ref{fig:T4Z2BlowdownLimit}) singularity, corresponding to the affine Dynkin diagram of $\E8$. In the blowdown procedure, the central torus node (blue) is blown down and one of the corner singularities (red) is blown up.}
 \label{fig:T4Z6BlowdownLimit}
\end{figure}
Fiberwise, this gives
\begin{itemize}
  \item one ``$\Z6$ fiber'' with the three fixed points, leading to an $\E8$ fiber after performing the blowup/blowdown procedure with vanishing orders $(4,5,10)$ (cf.\ figure \ref{fig:T4Z6BlowdownLimit}), 
  \item one ``$\Z3$ fiber'' which turns into an $\E6$ with vanishing orders $(3,4,8)$ as discussed in section \ref{sec:T4Z3DualConstruction}, and 
  \item one ``$\Z2$ fiber'', which becomes the $D_4$ fiber with vanishing orders $(2,3,6)$ as discussed in section \ref{sec:T4Z2DualConstruction}.
\end{itemize}
For the $\Z6$ fiber, this takes us from a finite component with multiplicity $6$ to a finite component with multiplicity $1$; the outcome of the procedure for the other fiber types are as discussed in the previous cases. We observe in general that in the description of the $\Z{N}$ singularities in a $T^4/\Z{N}$ orbifold, we obtain Dynkin diagrams in which the original orbifold corresponds to the ``central'' node (i.e.\ the node with the most neighboring nodes) which always has a Dynkin label of $N$.

The overall vanishing orders of $(\widetilde{f}_8,\widetilde{g}_{12},\widetilde{\Delta})$ would have to be $(9,12,24)$, which is impossible since $\widetilde{f}_8$ is a section in $8D_u$. Hence $\widetilde{f}_8=0$ identically, which again fixes $\tau=e^{2\pi\i/3}$ as required by the $\Z6$ orbifold action (albeit for a different reason than in the \Z3 case). We choose
\begin{align}
  d_{12}=\widetilde{g}_{12}=\alpha p_1^5 q_1^4 r_1^3\,,
\end{align}
 such that the $\Z6$, $\Z3$, and $\Z2$ fixed points are located at the zeros of $p_1$, $q_1$, and $r_1$, respectively.

The other polynomials are now relatively straightforward. By similar reasoning as above, we find
\begin{align}
  d_{24}=\beta p_1^{10} q_1^8 r_1^6\,.
\end{align}
Hence, the $\Z6$ fiber carries ten instantons. If we want to turn this into a fixed point instanton number $n_6$, we use that according to our previous results $n_3=\frac83$ and $n_2=\frac32$ and thus $n_6=10-\frac83-\frac32=\frac{35}{6}$, which amounts to $\frac76$ per $\P1$ glued in to resolve the $A_5$ singularity. This leads to an obvious pattern: A $\Z{k}$ fixed point carries instanton number
\begin{align}
 n_k=\frac{k^2-1}{k}\,.
\end{align}
The $\P1$'s glued into the corresponding $A_{k-1}$ singularity thus carry an instanton number of $\frac{k^2-1}{k(k-1)}=\frac{k+1}{k}$, if distributed evenly.

Finally, we discuss the charged matter. To fix $c_{20}$ we use again that each $\P1$ used to resolve a $\Z{N}$ singularity contributes one half-hyper $\rep{56}$. Hence we find
\begin{align}
  c_{20}&=\gamma p_1^8 q_1^6 r_1^4 k_2\,.
\end{align}
The distribution of the $\rep{56}$'s among the sectors and fixed points is then as follows:
\begin{itemize}
  \item The untwisted sector has one $\rep{56}$, corresponding to $k_2$,
  \item the $\theta$ sector has as its only fixed point the $\Z6$ fixed point at $p_1=0$, so we expect eight, minus the $\Z2$ and $\Z3$ fixed points in the $\Z6$ fiber which do not appear in the $\theta$ sector, minus the $\Z{2}$ and $\Z3$ components inside the $\Z6$, such that we end up with $8-1-2-1-2=2$ half-hypers, which can come as one charged multiplet,
  \item the $\theta^2$ sector contains the $\Z3$ fixed points at $q_1=0$, plus the ones in the $\Z6$ fiber, so $6+2+2=10$ half-hypers, again possibly charged,
  \item and finally the $\theta^3$ sector picks up the remaining $4+1+1=6$ half-hypers, which actually are true half-hypers, i.e.\ uncharged. 
\end{itemize}
This agrees perfectly with the orbifold spectrum given in table \ref{tab:OrbifoldSpectra}.

\subsubsection*{Stable degeneration}
Here we group the 24 $I_1$ fibers into a group with 10 $I_1$'s, a group with 8 $I_1$'s, and a group with $6$ $I_1$'s, resulting in one $\fibii^*$, one $\fibiii^*$, and one $I_0^*$ singularity intersecting $C_*$, which uses up all 24 $I_1$ fibers as in the previous cases. Again, we keep the $\E8$ fiber along $C_0$ and split the $\E8$ along $C_\infty$ into $\E7\times\U1$. When blowing up the central nodes, we obtain 1 $I_6$, 4 $I_3$, and 5 $I_2$ singularities, whose direct construction would require 28 $I_1$ fibers.

\subsection[Constructing the duals of \texorpdfstring{$T^4/\Z{N}$}{T4/Z{N}} orbifolds with Wilson lines]{Constructing the duals of \texorpdfstring{$\boldsymbol{T^4/\Z{N}}$}{T4/Z{N}} orbifolds with Wilson lines}
\label{sec:T4ZNWL}
As a last step we want to discuss orbifold models with Wilson lines. If on turns on Wilson lines at the orbifold, they further branch the gauge group and project out some of the matter states. As an example, we will discuss the $T^4/\Z2$ orbifold where one Wilson line $W=(0^2,1,0^5)(0^8)$ along one of the torus directions (say $e_1=\text{Re}(z_1)$) has been switched on in addition to the standard embedding shift vector $V$. As a result, the $2\times4=8$ $\Z2$ fixed points located at $e_{1}=0$ do not feel the Wilson line and behave as if it was not switched on, while the other $8$ $\Z2$ fixed points at $e_{1}=\frac12$ feel both the shift and the Wilson line. As a result, we get a gauge group of $\SO{12}\times\SU2\times\SU2'\times\E8$ with the matter content given in table \ref{tab:OrbifoldSpectra}. Note that the $\SU2$ corresponds to the factor branched off the $\E7$, while the $\SU2'$ corresponds to the ``original`` $\SU2$ of the above $\Z2$ example. At the $8$ fixed point where the Wilson line does not act (corresponding to the upper line in the third column of the $T^4/\Z2$ case in table \ref{tab:OrbifoldSpectra}), the spectrum is simply branched, which means that $(\rep{56},\rep{1})\rightarrow2(\rep{12},\rep{1})+(\rep{32},\rep{1})$. At the other $8$ fixed points, it has been calculated using the standard CFT techniques.

Hence the discussion is not as symmetric as in the first case. Nevertheless, the idea is similar. One starts with an $\SO{12}$ (i.e.\ $I_2^*$) singularity at $u=0$ with vanishing orders $(2,3,8)$ and an $\E8$ singularity at $v=0$ and factorizes the discriminant further by restricting the polynomials until two further $I_2$ loci appear which correspond to the two $\SU2$'s. The resulting expressions $f$, $g$, $\Delta$ then are of the form
\begin{subequations}
 \begin{align}
   f		&\propto u^2 v^4 p_4^2 \left(a_1 u^2+a_2p_4^3 uv+a_3p_4^6 v^2 \right)\,,\\
   g		&\propto u^3 v^5 \left(a_4 d_0u^4+\ldots+a_8p_4^{12}v^4 \right)\,,\\
   \Delta	&\propto u^8v^{10}\left(u+a_9 p_4^3 v \right)^2\left(u-a_9p_4^3 v \right)^2\left(a_{10}d_0u^2+\ldots+a_{12}p_4^{6}v^2\right)\,,
 \end{align} 
\end{subequations}
where the $a_i$ are (fixed) numerical constants that are determined such that the discriminant factorizes into two quadratic pieces plus a rest and that the terms containing $u^6$ and $u^7$ cancel from the discriminant, enhancing $D_4$ to $D_6$. We thus obtain an $\E8$ singularity at $v=0$, a $D_6$ singularity at $u=0$ and two $I_2$ singularities at $u\pm a_9p_4^3v=0$.

Applying the methods described in section \ref{sec:OtherConstructions} does again not give rise to the ``correct'' orbifold limit. Using table \ref{tab:GroupTheoryCoefficients}, we find
\begin{align}
\begin{split}
 (\alpha_{\SO{12}},\widetilde{\alpha}_{\SO{12}})=(2,12)\phantom{-}\,~~ &\quad\Rightarrow\quad \xi_{\SO{12}}  =D_v+12D_s=D_u\,,\\
 (\alpha_{\SU2},   \widetilde{\alpha}_{\SU2}) =(2,12)\phantom{-}\,~~   &\quad\Rightarrow\quad \xi_{\SU2}\;\;\!   =D_v+12D_s=D_u\,,\\
 (\alpha_{\SU2'},  \widetilde{\alpha}_{\SU2'})=(2,12)\phantom{-}\,~~   &\quad\Rightarrow\quad \xi_{\SU2'}~\! =D_v+12D_s=D_u\,,\\
 (\alpha_{\E8},    \widetilde{\alpha}_{\E8})  =(2,-12)\,~~  	       &\quad\Rightarrow\quad \xi_{\E8}~~~~\,  =D_v\,.
\end{split}
\end{align}
As expected, the $\SO{12}$ and the two $\SU2$'s are in the same divisor class $D_u$, and the $\E8$ is in the divisor class $D_v$. We thus find that 
\begin{align}
\begin{split}
 \xi_{\SO{12}}.\xi_{\E8} &=~~\;\xi_{\SU2}.\xi_{\E8}~~=~\;\xi_{\SU2'}.\xi_{\E8}~=0\,,\\
 \xi_{\SO{12}}.\xi_{\SU2}&=\xi_{\SO{12}}.\xi_{\SU2'}=\xi_{\SU2}.\xi_{\SU2'}=12\,.
\end{split}
\end{align}
Neglecting again the normalization such that fractional instantons can arise we find that $\xi_{\SO{12}}=\frac12D_u$, $\xi_{\SU2}=\xi_{\SU2'}=D_u$, $\xi_{\E8}=\frac{1}{60}D_v$, which gives the correct spectrum.

\subsection{General instanton embedding}
\label{sec:GeneralInstantonEmbeddings}
So far we have restricted our discussion to the case where all instantons are embedded in one $\E8$. In an orbifold with general gauge shift and Wilson lines, both $\E8$'s will be broken. Such models can also be connected to the ones we discussed here via a sequence of blowups and blowdowns in the base \cite{Seiberg:1996vs,Morrison:1996na,Morrison:1996pp}. By blowing up an extra point in the base, we can get from $\F{N}$ to $\F{N\pm1}$. The corresponding picture in the Ho\v{r}ava--Witten picture of M-theory \cite{Horava:1995qa,Horava:1996ma,Klemm:1996hh} is the following: the blowup in the base introduces an extra tensor multiplet; the scalar of this tensor multiplet encodes the position of on M5 brane in the M-theory bulk (i.e.\ on the interval between the two $\E8$ branes). Thus by blowing up and blowing down, we can ``peel off'' an M5 brane from one $\E8$ (blowup), let it travel through the bulk, and absorb it at the other $\E8$ brane (blowdown), changing the instanton embedding in this way. For the analysis, this means that the expansions in \eqref{eq:BasicExpansion} have to be done such that they are compatible with the new scalings. The geometry of the singularity structure does not change, and the terms in the expansions can still be identified with bundle and geometry data \cite{Morrison:1996pp}.


\section{Conclusions and outlook}
\label{sec:Conclusions}
In this paper we investigated heterotic models compactified on singular K3 spaces corresponding to abelian toroidal orbifolds $T^4/\Z{N}$ and argued that all these models can be connected via F-theory.

For the sake of simplicity we concentrated on the standard embedding of these orbifolds, in which one $\E8$ is broken rank-preservingly to either $\E7\times\SU2$ or $\E7 \times\U1$ and the other $\E8$ is left unbroken. First we discussed previous approaches for constructing F-theory duals of smooth heterotic K3 models and argued why these approaches cannot be used. For the construction of the duals, we first had to describe the geometry of the $T^4/\Z{N}$ models on the heterotic side. Having singular fibers by itself over singularities in the base, the description led to fibers with singularities of the type $I_0^*$, $\fibiv^*$, $\fibiii^*$ and $\fibii^*$. By calculating the $j(\tau)$ function of these fibers, we found that the complex structure of the orbifold torus is free in the $T^4/\Z2$ case, fixed to $\tau = e^{2\pi\i/3}$ in the $T^4/\Z{3}$ and $T^4/\Z{6}$ case, and fixed to $\tau=\i$  in the $T^4/\Z4$ case, as needed for compatibility with the orbifold action. We found that in these constructions the node corresponding to the ``orbifold pillow'' in a fiber above a $\Z{N}$ singularity in the base has multiplicity $N$. Furthermore, the distribution of the fractional instantons over the $\Z{N}$ fixed points in the base is $(N^2-1)/N $.

By blowing down the ``orbifold pillow'' and blowing up one of the Dynkin nodes with multiplicity one, we could go to a Weierstrass description of the model. Using the information about the instanton embeddings and the singularities of the geometry on the heterotic side, we could tune the complex structure parameters on the F-theory side such that the defining quantities $(f,g,\Delta)$ of the F-theory Weierstrass model have the vanishing orders needed to reproduce the gauge groups on the heterotic side. 

However, the resulting models were too singular to apply the usual spectrum calculation rules of F-theory. The reason is that (most) matter and instantons are living at the singularities on the heterotic side and thus all components of the discriminant on the F-theory side vanish at the fixed point loci of the heterotic dual. Nevertheless, by counting parameters and deforming the singularities away to connect to the (partially Higgsed) smooth model, we argued that the F-theory spectrum coincides with the spectrum on the heterotic side. Using the strong constraints imposed by 6D anomaly cancellation, we could further cross-check the spectrum. The results obtained from the counting agree with the ones obtained by Aspinwall and Donagi in \cite{Aspinwall:1998he} for F-theories with orbifold singularities in the fiber.

In the end, we investigated more general gauge sectors with Wilson lines and configurations in which both $\E8$'s were broken on the heterotic side and argued that these can be obtained by further specialization of the complex structure moduli in the CY threefold combined with blowups and blowdowns in the base $\F{N}$. In particular, the geometry on the heterotic side and the gauge group and matter content in each $\E8$ are described by the complex structure on the F-theory side, while the instanton embedding on the heterotic side corresponds to blowups and blowdowns in the base on the F-theory side. This means that heterotic orbifold Wilson lines map to complex structure parameters (and not to Wilson lines) on the F-theory side.

\subsection*{Outlook}

It should be noted that the methods we applied here for the description of the geometry of the singular K3 compactifications could in principle also be applied to CY threefolds on the heterotic side (of course there the anomaly cancellation conditions are weaker, which makes cross-checking the F-theory matter spectrum harder). It would be very interesting to see which of the $T^6/\Z{N}$ and $T^6/(\Z{N}\times\Z{M})$ orbifolds are connected via F-theory on CY fourfolds. 

In addition, one could work out the description of the heterotic orbifolds and their F-theory duals directly using multisections as recently explored in \cite{Braun:2014oya} rather than constructing the Weierstrass model using the blowup and blowdown procedure, and match the results.

Furthermore, it would be interesting to study the M-theory description of these models and compare it with the the case of frozen singularities and discrete 3-form flux \cite{Witten:1997bs,deBoer:2001px}.

\subsection*{Acknowledgments}
We thank Andreas Braun, Volker Braun, and I\~naki Garc\'{\ii}a Etxebarria for helpful discussions and Denis Klevers for comments on the manuscript. The work of FR was supported by the German Science Foundation (DFG) within the Collaborative Research Center (SFB) 676 ``Particles, Strings and the Early Universe''.

\begin{footnotesize}
\providecommand{\href}[2]{#2}\end{footnotesize}
\end{document}